\PassOptionsToPackage{table}{xcolor}
\documentclass[]{fairmeta}

\DeclareTextFontCommand{\textbf}{\bfseries}

\usepackage{amsmath,amsfonts,bm}

\def\eqref#1{equation~\ref{#1}}

\def\1{\bm{1}}

\DeclareMathAlphabet{\mathsfit}{\encodingdefault}{\sfdefault}{m}{sl}
\SetMathAlphabet{\mathsfit}{bold}{\encodingdefault}{\sfdefault}{bx}{n}

\usepackage{amssymb}
\usepackage{amsthm}
\usepackage{tikz}
\usepackage{longtable}
\usepackage{adjustbox}
\usepackage{pifont}
\usepackage{colortbl}
\usepackage{tabularx}
\usepackage{array}
\newcommand{\TOCdots}{\nobreak\dotfill\nobreak}

\newcounter{findingctr}
\newtcolorbox{findingbox}{
  enhanced, sharp corners,
  colback=gray!9, colframe=black!65,
  boxrule=0pt, leftrule=2.5pt,            
  left=6pt, right=6pt, top=3pt, bottom=3pt,
  before skip=6pt, after skip=4pt,
  fontupper=\small
}
\newcommand{\finding}[2]{%
  \refstepcounter{findingctr}\label{#1}%
  \begin{findingbox}\textbf{Finding~\thefindingctr:}~#2\end{findingbox}}

\definecolor{TaskGuidance}{HTML}{315D92}
\definecolor{TaskPlanning}{HTML}{2D7161}
\definecolor{TaskAccessibility}{HTML}{725398}
\definecolor{TaskDecision}{HTML}{92611E}
\definecolor{TaskPrevention}{HTML}{A54F50}
\definecolor{TaskFinding}{HTML}{327783}

\definecolor{BenchBlue}{HTML}{1F4E79}
\definecolor{BenchLight}{HTML}{EEF5FA}
 
\DeclareUrlCommand\bcode{\urlstyle{tt}}

\makeatletter
\newcommand{\btask}[1]{\texttt{\btask@split#1-\@nil}}
\def\btask@split#1-#2\@nil{#1\if\relax\detokenize{#2}\relax\else
  -\allowbreak\btask@strip#2\fi}
\def\btask@strip#1-{#1}
\makeatother
 
\newcommand{\bTaskHead}[1]{%
  \toprule
  \textbf{Task} & \textbf{#1} & \textbf{Evaluator validation} \\
  \midrule
  \endfirsthead
  \toprule
  \textbf{Task} & \textbf{#1} & \textbf{Evaluator validation} \\
  \midrule
  \endhead
  \bottomrule
  \endfoot}
 
\newtcolorbox{bprompt}[1]{
  enhanced,
  breakable,
  colback=BenchLight,
  colframe=BenchBlue,
  boxrule=0.55pt,
  arc=1.2mm,
  left=1.5mm,right=1.5mm,top=1.2mm,bottom=1.2mm,
  title={#1},
  fonttitle=\bfseries,
  fontupper=\small
}

\definecolor{memorygray}{RGB}{240,235,248}
 
\title{InteractiveWorld: A Verifiable Human-Centric Environment for Building Better Interactive MLLM Assistants}

\author[1,2,*]{Qinsi Wang}
\author[1]{Wei Wen}
\author[1]{Chong Zhou}
\author[1]{Saksham Suri}
\author[1]{Zechun Liu}
\author[1]{Lemeng Wu}
\author[2]{Hai Helen Li}
\author[2]{Yiran Chen}
\author[1]{Raghuraman Krishnamoorthi}
\author[1]{Qi Qian}

\affiliation[1]{Meta}
\affiliation[2]{Duke University}

\contribution[*]{Work completed during an internship at Meta.}

\abstract{People have long dreamed of an assistant that is always there to guide them and remind what they forget, from new dishes to lost keys. Powered by multimodal large language models (MLLMs), egocentric AI assistants on wearable devices are bringing this dream within reach: they see what the user sees and give instructions for user's physical work when necessary.
However, testing such assistance is challenging and usually requires real people. The expensive and rare real interaction data leaves a key question unanswered: \textbf{\textit{Can today's MLLMs truly help people in physical world?}}
To answer this question, we propose \textbf{InteractiveWorld}, the first verifiable environment for closed-loop interaction between human and AI assistant. Unlike existing robot-centric simulators, InteractiveWorld lets the assistant guide a human agent who observes the scene, understands instructions, and acts independently. It thus realistically simulates how AI assistant collaborates with human: the assistant instructs, the human acts, and the environment executes actions and updates what they see next. Building on it, we introduce \textbf{InteractiveBench}, the first dynamic benchmark for egocentric AI assistants, with 120 multi-turn tasks across six categories. In each task, the assistant guides the human in real time, and every instruction shapes the outcome, which is scored by verifiable, state-based rewards. Experiments with 12 frontier MLLMs show they are still far from reliable assistants. Even the strongest model, Claude Fable 5.1, achieves at most 20\% and 25\% on long-horizon Guidance and Planning tasks, respectively. Overall, our work offers a verifiable platform and a rigorous evaluation set, laying a solid foundation for research on interactive MLLM assistants that truly help people in the physical world.}

\date{\today}
\correspondence{\email{qiqian@meta.com}}

\metadata[Code]{\url{https://github.com/facebookresearch/InteractiveWorld}}
\metadata[Project Page]{\url{https://meta-interactiveworld.github.io/}}
\begin{document}

\let\cite\citep

\maketitle
 
\section{Introduction}
A long-standing goal of AI is to help people in the physical world. In everyday tasks, people need a companion that sees what they see, remembers what they forget, and tells them what to do next. \textbf{\textit{Egocentric AI assistants}} aim to fill this role: running on wearable devices, they share the user's first-person view and guide users in natural language. Unlike robots, they avoid physical control: the human acts, while the assistant focuses on perception, memory, reasoning, and communication. As MLLMs rapidly improve~\citep{anthropic2026fable51,openai2026gpt56}, it makes egocentric AI assistants a practical, near-term path to helping people in the physical world.

However, evaluating whether assistants truly help people remains challenging. Real-world experiments are costly and hard to scale with human. Moreover, existing benchmarks are almost all \textit{\textbf{static}}: a model watches a pre-recorded first-person video and answers questions about it~\citep{grauman2022ego4d,cheng2024egothink} or predicts when and how to intervene~\citep{wang2023holoassist,zhang2025proassist}. Whatever the model says, the video remains unchanged.
On the contrary, real-world assistance is \textit{\textbf{dynamic}}: the assistant and human interact over many turns with diverse rewards, requiring abilities that fixed videos cannot test.
\textbf{(1) Closed-loop adaptation.} The assistant must continually adapt its guidance to the evolving scene and human responses, since its instructions directly change the human's actions and what it sees next.
\textbf{(2) Compositional capabilities.} The assistant must combine skills such as tracking object states, remembering events, and communicating clearly, whereas static benchmarks typically test one skill at a time.
This leads to core question: \textbf{\textit{How can we faithfully evaluate an assistant's ability to interact with and help humans in real world?}}

To answer this question, we propose \textbf{InteractiveWorld}, to the best of our knowledge the first environment for closed-loop interaction between humans and assistants. Built on Isaac Sim~\citep{nvidia2024isaacsim}, it contains over 8,000 objects and 50 scenes. As shown in Fig.~\ref{fig_main} (top), interactive loop in InteractiveWorld consists of three components: 
(1) an \textit{Assistant agent} that, like a real wearable assistant, shares the human's egocentric view, receives the task from the human, and decides when and how to offer natural-language guidance;
(2) a \textit{Human agent}, driven by an MLLM, that interprets the guidance based on its observations and converts it into high-level commands from 36 action primitives (e.g., \texttt{[Grab]<cup>}), much as a person translates advice into behavior; and
(3) an \textit{Interactive Environment} that stably executes these commands through a full-body human model capable of autonomous navigation, posture adaptation, and collision-free manipulation, while continuously rendering the resulting first-person view.
Through this closed loop, \textbf{\textit{InteractiveWorld realistically simulates how an assistant helps a person with everyday tasks}}.

Building on InteractiveWorld, we introduce \textbf{InteractiveBench}, the first closed-loop dynamic benchmark for egocentric AI assistants, comprising 120 human-reviewed tasks across six categories: \textit{Guidance, Planning, Accessibility, Decision, Prevention, and Finding}. Each task requires the assistant to guide the human over many turns by combining perception, memory, reasoning, and communication.
Task success is verified automatically from the final object states in the simulator.
Since the assistant's guidance directly shapes the human's actions for the task, InteractiveBench can faithfully exam whether a model helps human. \textit{We also provide several demos in the supplementary material.}

Experiments with 12 frontier MLLMs on InteractiveBench show \textbf{\textit{state-of-the-art models still have substantial room for improvement in interactive assistance.}}
Even the strongest model, Claude Fable 5.1, succeeds on only 41.7\% of the tasks, and all other models stay at or below 33.3\%. In particular, on Guidance and Planning, the best models reach only 20.0\% and 25.0\%, respectively, indicating models still perform poorly on long-horizon interactive tasks. We further analyze model behaviors in detail and distill five key findings in Sec.~\ref{sec:main_results}. In summary, our contributions are as follows:

\begin{figure*}
 \centering
 \centerline{\includegraphics[width=\linewidth]{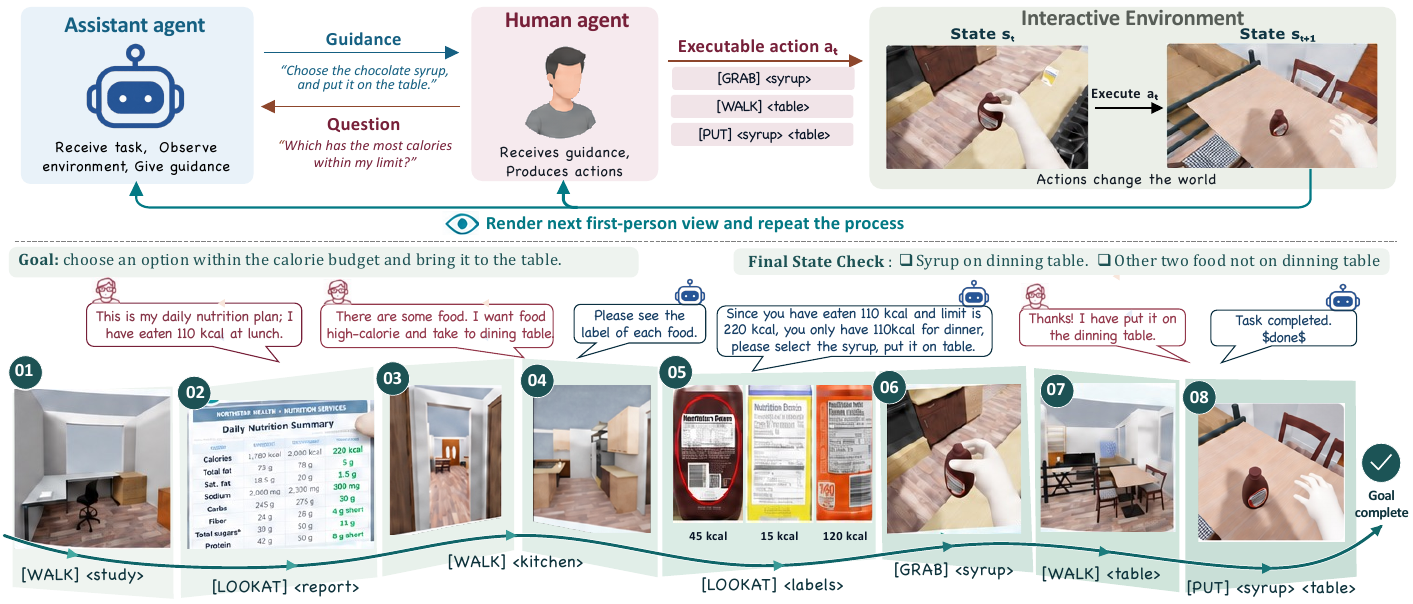}}
 \caption{(Top) \textbf{Interaction Loop in InteractiveWorld.} The Assistant agent provides guidance based on current observations and history, which the Human agent grounds into executable actions to update the environment.(Bottom) \textbf{Example task from InteractiveBench.} The Assistant helps the human choose dinner (using a daily plan and food labels) and guides its placement on the table. Interaction relies on natural language and is evaluated by state-verifiable metrics.}
 \label{fig_main}
 \vspace{-5pt}
\end{figure*}

\begin{itemize}
    \item We present \textbf{InteractiveWorld}, the first verifiable human-centric interaction environment, in which an assistant's natural-language guidance drives a human agent's actions and changes the world, enabling realistic simulation of their everyday collaboration.
    \item We introduce \textbf{InteractiveBench}, the first multi-turn, closed-loop benchmark for evaluating models as egocentric AI assistants in the physical world, where assistant outputs shape task outcomes and success is measured by verifiable, state-based rewards.
    \item Experiments reveal major gaps in current MLLMs' dynamic interaction. Our analyses identify perception, memory, and communication as key bottlenecks. \textit{These findings underpin future interactive MLLM research, with InteractiveWorld providing a verifiable platform.}
\end{itemize}



\section{InteractiveWorld}
In this section, we introduce InteractiveWorld, the first closed-loop human–assistant interaction environment providing a realistic, verifiable platform for evaluating egocentric AI assistants.

\textbf{Challenges}.
Existing simulators provide robust physics and rich 3D assets for embodied AI. However, they are robot-centric, focusing on low-level motion planning and direct joint control. Transitioning to egocentric AI assistants introduces a paradigm shift: \textbf{\textit{the agent must guide human via natural language rather than directly manipulating hardware.}} Simulating interaction between human and assistant requires modeling the human as a capable collaborator with three key aspects:
\begin{itemize}
\item \textbf{Human-centric egocentric view.} The assistant observes the world through the human's first-person view, which constantly shifts with body postures and hand movements.
\item \textbf{Reliable execution of everyday actions.} Unlike robots, humans can reliably and naturally execute everyday actions (e.g., \texttt{GRAB}, \texttt{OPEN}) given high-level intents.
\item \textbf{Autonomous translation of guidance into actions.} Humans can autonomously translate free-form guidance into executable actions based on what they see in the current scene.
\end{itemize}
Thus, while existing simulators provide a foundation, building a human-centric environment requires a new human representation, reliable semantic execution, and interaction mechanism.

\begin{figure*}
 \centering

 \centerline{\includegraphics[width=0.93\linewidth]{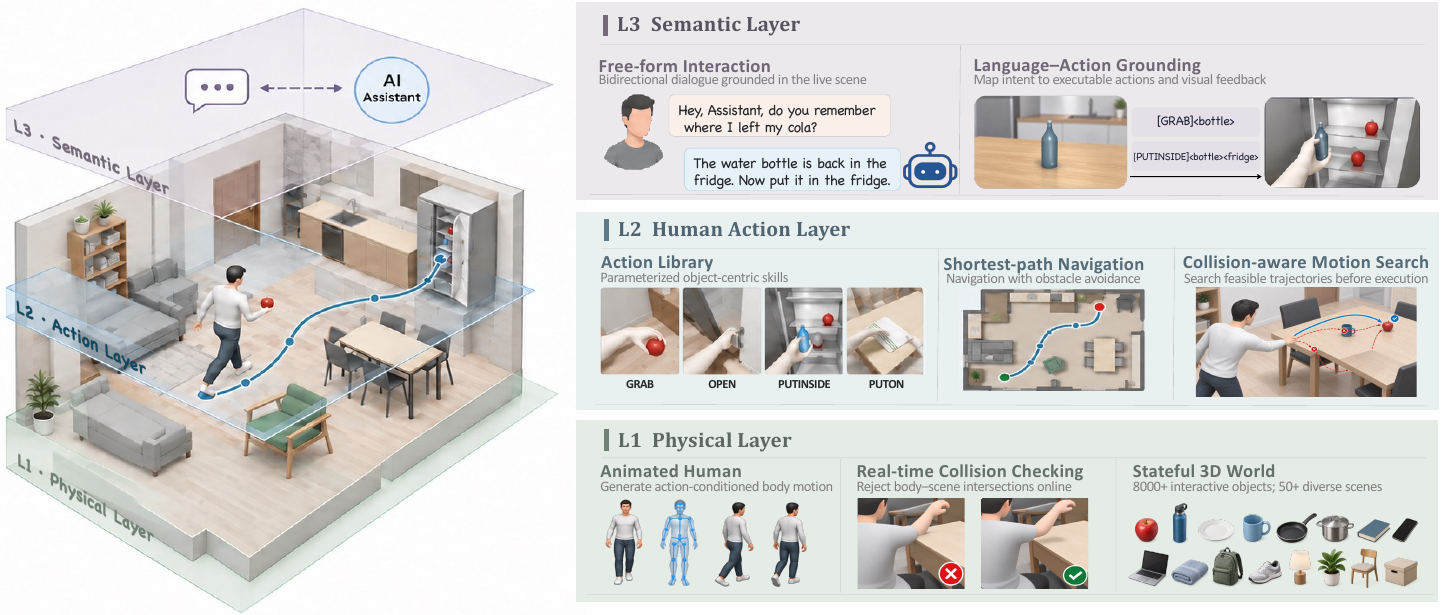}}
 \caption{\textbf{InteractiveWorld Overview.} Three layers enable realistic human--assistant collaboration: (1) \textit{Physical}: realistic humans, assets, and physics simulation; (2) \textit{Human Action}: autonomous, stable high-level daily action execution; (3) \textit{Semantic}: grounding free-form guidance into action sequences.}
 \label{fig_layer}
\end{figure*}

\textbf{Overall.} To address these challenges, we propose InteractiveWorld, a human-centric interaction environment built on Isaac Sim (Fig.~\ref{fig_layer}). From bottom to top, it consists of three layers: (1) The Physical Layer ensures stable, repeatable human actions that are physically consistent with the environment. (2) The Human Action Layer enables the ``Human'' to navigate and manipulate objects naturally and autonomously. (3) The Semantic Layer connects natural language guidance to executable human behaviors. Together, these layers make the Human \textbf{\textit{an interactive agent that observes the environment, understands instructions, and executes actions independently}}, rather than merely a directly controlled animated character. This enables InteractiveWorld to realistically simulate collaboration between human and assistant. We introduce each module in detail below.

\subsection{Physical Layer}
The physical layer serves as a foundational 3D environment, providing realistic humans, assets, and physics engine. It supplies the basic physical states needed for high-level planning.

\textbf{Stable Human Representation.} We use SMPL+H ~\citep{romero2017embodied} to model the full body and hands, placing a camera on the head for a first-person visual perspective. To prevent physical failures like random falls, the human model is driven kinematically. This ensures that the physical layer can successfully and deterministically execute any valid pose sequence provided by the upper layer.

\textbf{Collision-Free Interaction.} To prevent severe collisions between the human model and the environment, the system pre-checks navigation paths and arm reachability before execution. It automatically corrects minor collisions by smoothly adjusting the grasp position. For unsafe actions, the system rejects them and returns clear failure reasons to the upper layer.

\textbf{Rich and scalable asset support.} To support diverse daily tasks, we imported over 8,000 object and more than 50 environment assets from BEHAVIOR-1K, including furniture, containers, appliances, tools, and food. Thanks to Isaac Sim's asset import features, InteractiveWorld can load scanned models of real-world scenes and objects to simulate specific environments for future development.


\subsection{Human Action Layer}
Built on the physical foundation, the Human Action Layer enables the human model to perform high-level actions (e.g., grab, open) by dividing the execution into two stages:

\textbf{Collision-Aware Auto-Navigation.} Given a target object, the system first identifies a reachable standing position suitable for manipulation. It then constructs an obstacle map based on environment and plans a safe path using A* search \citep{hart1968formal}. Following this path, the human executes continuous walking and turning motions, moving smoothly toward target without collisions.


\textbf{Target-Driven Adaptive Manipulation.} Upon arrival, the system adapts its posture to stand, lean, or squat and plans a collision-free arm trajectory. Grasp position and hand posture depend on object shape, functional parts, and task. Predefined motion sequences for basic postures and actions let the hand wrap around a cup to lift it or grasp a handle to open a refrigerator. If the initial position offers no feasible trajectory to the object, the system replans its position.


\subsection{Semantic Layer}

The Semantic layer builds upon the Human Action layer to link natural language with behavior, enabling the Assistant to guide the Human via dialogue rather than direct simulator control.


\textbf{Unified action interface.} We define 36 executable action primitives covering navigation, observation, manipulation, and daily activities. Supported actions range from basic (\texttt{[Walk]}, \texttt{[LookAt]}, \texttt{[Grab]}, \texttt{[Put]}, \texttt{[PutInside]}, \texttt{[Open]} and \texttt{[TurnOn]}) to complex operations (\texttt{[Pour]}, \texttt{[Cook]}, \texttt{[Cut]} ). Commands specify only actions and targets (e.g., \texttt{[Grab]<cup>} or \texttt{[PutInside]<pizza> <oven>}) and are translated into goals for the human action layer, which autonomously handles underlying navigation, positioning, and body motions.

\textbf{Verifiable multi-turn interaction.} During interaction, the Assistant provides natural-language guidance based on the Human’s first-person view and the conversation history. The Human interprets this guidance, then selects and executes the corresponding actions. After each action, the simulator updates the environment state and renders a new first-person view, forming a continuous "observe-guide-execute" loop. By tracking actual object positions and states throughout the interaction, the system can evaluates final object relationships to automatically verify task success.

By integrating these three layers, InteractiveWorld builds a comprehensive pipeline from the physical environment and human actions to natural language interaction. With an appropriate Human agent and Assistant agent, the system can simulates real-world scenarios where an egocentric AI Assistant assists a human, and provide verifiable outcomes to evaluate the Assistant's helpfulness.

\begin{table*}[t]
\centering
\caption{
\textbf{Overview of the InteractiveBench Classification.} To simulate real-world scenarios, visual inputs are intentionally withheld from human agents during accessibility tasks. 
}
\label{tab:task_overview}
\vspace{-8pt}
\begingroup
\fontsize{8.5}{10}\selectfont
\setlength{\tabcolsep}{5pt}
\setlength{\aboverulesep}{1.2pt}
\setlength{\belowrulesep}{1.2pt}
\renewcommand{\arraystretch}{1.02}

\begin{tabularx}{\textwidth}{@{}
  >{\raggedright\arraybackslash}p{0.12\textwidth}
  >{\raggedright\arraybackslash}p{0.255\textwidth}
  >{\raggedright\arraybackslash}X
@{}}

\toprule[0.8pt]
\textbf{Category} &
\textbf{Evaluation focus} &
\textbf{Example task} \\
\midrule[0.5pt]

\textcolor{TaskGuidance}{\textbf{Guidance}}
&
\textcolor{TaskGuidance}{\textbf{Guide user step by step}}
based on the user's progress, helping them complete task.
&
\textcolor{TaskGuidance}{\textbf{Lab cleanup.}}
A new lab worker must put away 14 chemical bottles and
clear broken glass. The assistant must use the labels, chemical knowledge
and the safety rules to guide the cleanup.
\\
\midrule[0.25pt]

\textcolor{TaskPlanning}{\textbf{Planning}}
&
\textcolor{TaskPlanning}{\textbf{Outline the steps}}
for tasks within time and resource limits, flexibly adjust plan.
&
\textcolor{TaskPlanning}{\textbf{Cooking for guests.}}
A chef need cook for guests with different food allergies,
using one oven and pan within 30 minutes.
The assistant must choose and schedule as many safe dishes
as possible.
\\
\midrule[0.25pt]

\textcolor{TaskAccessibility}{\textbf{Accessibility}}
&
\textcolor{TaskAccessibility}{\textbf{Describe the scene for blind}}
or low-vision users so they can
complete everyday tasks.
&
\textcolor{TaskAccessibility}{\textbf{Kitchen cleanup.}}
A blind user needs to clean a kitchen with broken glass,
food, and dishes. The assistant must read the cleanup guide,
describe the scene, and help the user
finish safely.
\\
\midrule[0.25pt]

\textcolor{TaskDecision}{\textbf{Decision}}
&
\textcolor{TaskDecision}{\textbf{Help with daily decision}}
using current observations and assistant's knowledge.
&
\textcolor{TaskDecision}{\textbf{Supplier comparison.}}
A buyer sees an order for 40 filters,
then sees two quotes with different box sizes in another room.
The assistant must choose the
cheaper offer for the full order.
\\
\midrule[0.25pt]

\textcolor{TaskPrevention}{\textbf{Prevention}}
&
\textcolor{TaskPrevention}{\textbf{Warn before mistakes happen}},
while avoiding unnecessary interruptions.
&
\textcolor{TaskPrevention}{\textbf{Leaving home.}}
A user is about to leave, but the fridge and living-room
window were left open earlier. The assistant must remember
these problems and remind the user before they leave.
\\
\midrule[0.25pt]

\textcolor{TaskFinding}{\textbf{Finding}}
&
\textcolor{TaskFinding}{\textbf{Help find missing objects}}
by remembering where they moved and guiding search.
&
\textcolor{TaskFinding}{\textbf{Finding stored tools.}}
A user cannot find the black and blue squeegees after
storing them in a lidded box inside a cabinet in another room.
The assistant must remember where they went.
\\

\bottomrule[0.8pt]
\end{tabularx}

\endgroup
\end{table*}

\section{InteractiveBench}

Building on InteractiveWorld, we further propose InteractiveBench. To the best of our knowledge, this is the first closed-loop interactive benchmark designed for egocentric AI assistants.


\textbf{Task Design.} InteractiveBench contains 120 tasks across six categories (Tab.~\ref{tab:task_overview}): Guidance, Planning, Accessibility, Decision, Prevention, and Finding. These cover the main everyday scenarios for egocentric AI assistance.
All tasks undergo human review and meet three criteria:
First, each addresses a realistic need for assistance, rather than a simple action the Human can complete unaided.
Second, it requires the Assistant to use first-person information to observe, remember, reason, or plan, and to help over multiple turns.
Third, outcomes must be verifiable from clearly defined environment states. Appendix~\ref{app:interactivebench} details the full task construction and review process.



\definecolor{rankred}{RGB}{200,40,40}
\definecolor{rankblue}{RGB}{40,90,180}
\definecolor{rankgreen}{RGB}{40,140,70}

\newcommand{\first}[1]{\textcolor{rankred}{\textbf{#1}}}
\newcommand{\second}[1]{\textcolor{rankblue}{\textbf{#1}}}
\newcommand{\third}[1]{\textcolor{rankgreen}{\textbf{#1}}}

\begin{table*}[!t]
    \centering
    \caption{
    \textbf{Performance Across Six Task Categories.}
    Pass is the percentage of fully completed instances (all objectives met).
    MCR (Mean Completion Rate) is the average fraction of objectives completed per instance.
    We verified that, under our evaluation protocol, the same Human agent achieves a 100\% pass rate with an Oracle Assistant and 0\% without assistance.
    }
    \label{tab:task_performance}
    \label{table_pass_rate}
    \renewcommand{\arraystretch}{1.1}   
    \setlength{\tabcolsep}{4pt}
    \resizebox{0.95\textwidth}{!}{%
    \begin{tabular}{l cc cc cc cc cc cc cc}
        \toprule[2pt]

       \multirow{2}{*}{\textbf{Model}}
& \multicolumn{2}{c}{\textbf{AVG.}}
& \multicolumn{2}{c}{\textbf{Guidance}}
& \multicolumn{2}{c}{\textbf{Planning}}
& \multicolumn{2}{c}{\textbf{Accessibility}}
& \multicolumn{2}{c}{\textbf{Decision}}
& \multicolumn{2}{c}{\textbf{Prevention}}
& \multicolumn{2}{c}{\textbf{Finding}} \\

        \cmidrule(lr){2-3}
        \cmidrule(lr){4-5}
        \cmidrule(lr){6-7}
        \cmidrule(lr){8-9}
        \cmidrule(lr){10-11}
        \cmidrule(lr){12-13}
        \cmidrule(lr){14-15}

        & \shortstack{\textbf{Pass}\\(\%)}
        & \shortstack{\textbf{MCR}\\(\%)}

        & \shortstack{\textbf{Pass}\\(\%)}
        & \shortstack{\textbf{MCR}\\(\%)}

        & \shortstack{\textbf{Pass}\\(\%)}
        & \shortstack{\textbf{MCR}\\(\%)}

        & \shortstack{\textbf{Pass}\\(\%)}
        & \shortstack{\textbf{MCR}\\(\%)}

        & \shortstack{\textbf{Pass}\\(\%)}
        & \shortstack{\textbf{MCR}\\(\%)}

        & \shortstack{\textbf{Pass}\\(\%)}
        & \shortstack{\textbf{MCR}\\(\%)}

        & \shortstack{\textbf{Pass}\\(\%)}
        & \shortstack{\textbf{MCR}\\(\%)} \\

        \midrule

        Claude Fable 5.1
& \first{41.7} & \first{54.8}
& \first{20.0} & \second{32.3}
& \first{25.0} & \second{45.6}
& 30.0 &  63.3
& \first{85.0} & \first{95.1}
& \first{50.0} & \first{50.0}
& \first{40.0} & \first{42.7} \\

Claude Opus 5
& \second{33.3} & \second{48.1}
& 15.0 & \first{33.7}
& \first{25.0} & \first{47.3}
& 25.0 & 56.5
& \second{70.0} &  86.4
& 30.0 & 30.0
& \second{35.0} &  35.0 \\

Muse Spark 1.1
&  31.7 &  43.7
& \first{20.0} & 28.8
& \first{25.0} & 34.6
&  35.0 & 63.0
& 55.0 & 76.9
& 30.0 & 30.0
& 25.0 & 29.0 \\

GPT-5.6 Sol
& 30.8 & 42.0
& \first{20.0} & 31.3
& \first{25.0} & 36.4
& 30.0 & 54.8
& \second{70.0} & \second{89.4}
& 15.0 & 15.0
& 25.0 & 25.0 \\

GPT-5.5
& 30.0 & 41.2
& 10.0 & 20.3
& 15.0 & 32.8
& \first{45.0} & 64.2
& 65.0 & 82.1
& 25.0 & 25.0
& 20.0 & 22.7 \\

GLM 5.3 Flash
&30.0   &  43.1
&15.0   &22.4  
&20.0  & 42.4
&50.0 & 73.0 
&50.0   & 76.7 
&20.0   & 21.1 
&25.0 & 25.0 \\

Claude Opus 4.8
& 28.3 & 41.7
& 0.0 & 15.6
& 20.0 & 41.0
& 30.0 & 49.5
& 50.0 & 72.4
& \second{35.0} & \second{35.0}
& \second{35.0} & \second{36.8} \\

Kimi K3
&24.2   &  41.1
&\first{20.0}   &  28.0
&15.0  & 39.1 
&\second{40.0} &  \second{65.8}
&35.0   &  76.1
&20.0   &  21.1
&15.0 &  19.0\\

Qwen 3.8 Max
&22.5   & 36.0 
&5.0   &  9.7
&20.0  &  33.4
& 35.0 & 61.0 
&35.0   &  71.6
&20.0   &  20.0
&20.0 &  24.0\\

Gemini 3.1 Pro
& 20.8 & 37.8
& 10.0 & 22.7
& 10.0 & 37.5
& 35.0 & \first{65.9}
& 50.0 & 79.1
& 10.0 & 10.0
& 10.0 & 11.8 \\

GPT-5.6 Luna
& 20.0 & 37.2
& 5.0 & 22.0
& 5.0 & 28.7
& 20.0 & 51.3
& 40.0 & 66.2
& 30.0 & 30.0
& 20.0 & 24.9 \\

DeepSeek V4.1 Flash
&12.5   &  28.2
&0.0   &  11.4
&5.0  &  24.0
&25.0&  56.9
&35.0  &  71.0
&0.0   & 0.0 
&10.0 &  10.0\\

        \bottomrule[2pt]
    \end{tabular}%
    }

    \vspace{8pt}  

    \includegraphics[width=0.95\linewidth]{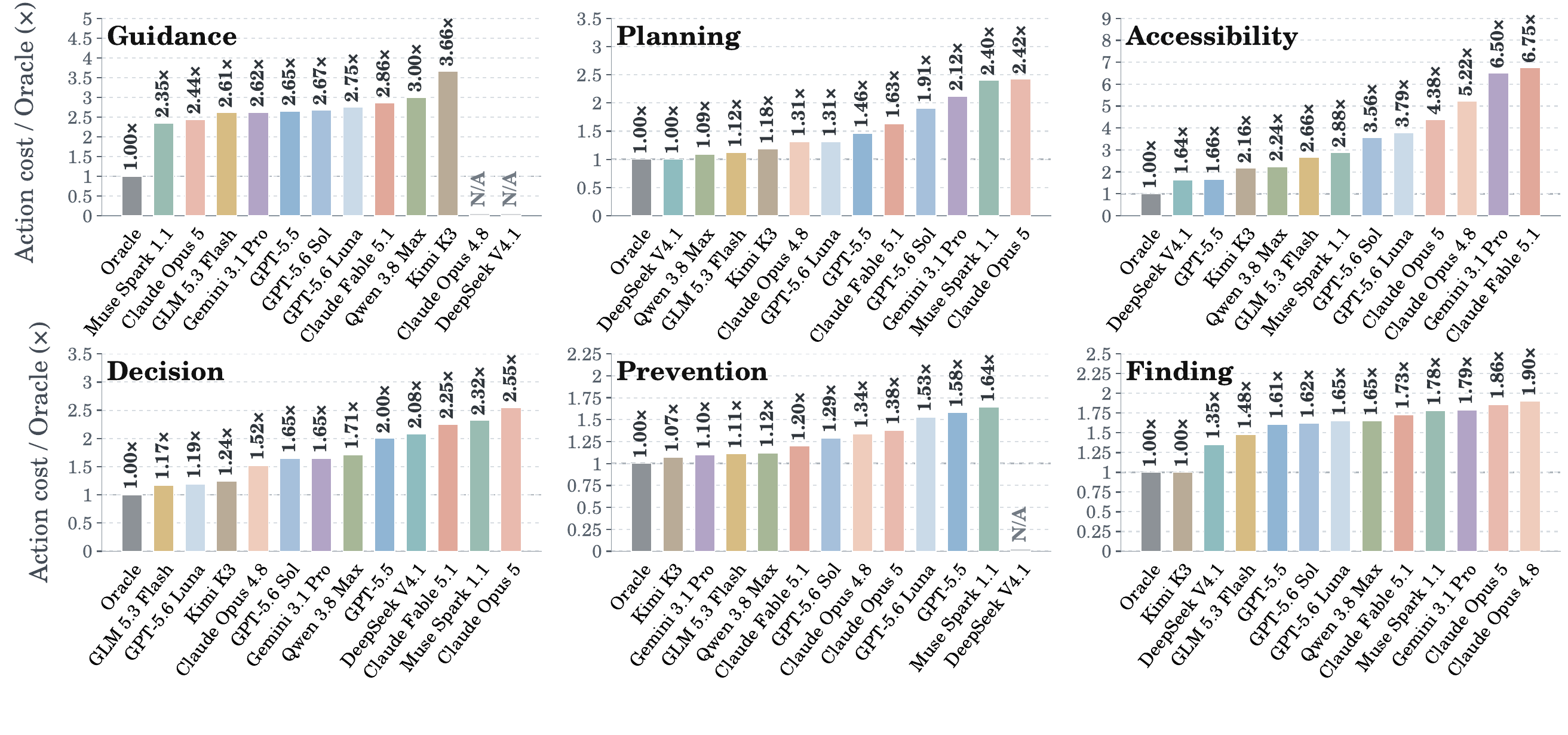}
    \vspace{-15pt}
    \captionof{figure}{\textbf{Action Efficiency of Evaluated Models.} Efficiency is measured by the average ratio of a model's action steps to the task's oracle steps (with oracle assistant) on its individually passed tasks.}
    \label{fig_efficiency}
\end{table*}

\textbf{Task Evaluation.} During evaluation, the target model serves as the Assistant agent and an auxiliary one acts as the Human agent. The Human agent is provided with all valid commands and can freely perform actions. However, it is instructed to follow the Assistant's guidance rather than acting on its own intentions. Meanwhile, the Assistant agent is prompted to help the human complete the task. Detailed settings are in Appendix \ref{app:interactivebench}. Overall, each evaluation consists:
\begin{itemize}
    \item \textbf{Initial Scene.} Each task defines a complete starting world. This includes the scene layout, the human's starting position, and the locations and states of target and distractor objects.
    \item \textbf{Pre-interaction Setup (Warmup).} Before formal interaction begins, the human performs predefined actions and provides necessary opening context (e.g., looking around or reading). This stage establishes the background and gives assistant visual and temporal information.
    \item \textbf{Closed-loop Interaction.} After warmup, the assistant and human begin interaction. The assistant uses the egocentric view and conversation history to choose whether to stay silent, provide guidance, or end the task. The human then issues action commands and feedback based on the assistant's guidance and their own observations. The action executor performs these commands in the environment. After each action, both sides get a new chance to respond, and keyframes sampled during the action are added to the visual input.
    \item \textbf{Termination conditions.} The task ends and evaluation begins when the assistant outputs \$done\$. Alternatively, it force-terminates upon reaching a maximum limit of interaction.
    \item \textbf{Task Scoring.} Each task has a predefined set of core states (e.g., "the pizza is in the oven" or "the fridge is closed"). When the task ends, the system checks the actual states of the environment. The task is marked successful only if all required states are met.
\end{itemize}

\textbf{Key advantages.} InteractiveBench measures how the Assistant’s advice affects subsequent actions and outcomes, offering three advantages:
(1) \textit{Long-horizon closed-loop interaction.} Each task requires multiple rounds of observation, guidance, action, and feedback.
(2) \textit{Multi-Skill integration.} Tasks require combining visual perception, memory, reasoning, and communication.
(3) \textit{Verifiable outcomes.} Success is determined by actual object states in the simulator.
By simulating realistic everyday interactions, InteractiveBench measures whether the Assistant can help the Human.

\section{Evaluation}
\label{sec:experiments}
In this section, we evaluate the performance of frontier models as the Assistant agent on the proposed InteractiveBench to demonstrate the current challenge of MLLMs for real scenarios.

\textbf{Models.}
We benchmark twelve state-of-the-art models as the assistant: Claude Fable 5.1, Claude Opus 5, and Claude Opus 4.8 from Anthropic \citep{anthropic2026fable51,anthropic2026opus5,anthropic2026opus48}; GPT-5.6 Sol, GPT-5.6 Luna, and GPT-5.5 from OpenAI \citep{openai2026gpt56,openai2026gpt55}; Muse Spark 1.1 from Meta \citep{meta2026musespark11}; GLM-5.3-Flash from Z.ai \citep{glm5team2026glm5}; Kimi K3 from Moonshot AI \citep{moonshot2026kimik3}; Qwen3.8-Max from Alibaba \citep{alibaba2026qwen38max}; Gemini 3.1 Pro from Google \citep{google2026gemini31pro}; and DeepSeek-V4.1-Flash from DeepSeek \citep{deepseek2026v41flash}.

\textbf{Evaluation Setup.} To ensure consistent Human behavior and fair Assistant comparison, our main experiments keep Gemini 3.1 Pro as the Human agent and Section 4.2 ablates the impact of varying Human agents. Visual inputs are capped at 32 frames per turn, sampled at 4 frames per action. If this limit is exceeded, we retain the 8 most recent frames and uniformly sample 24 from the older history. Per turn, the Assistant outputs \texttt{$interrupt$} (with natural language instructions), \texttt{$silent$}, or \texttt{$done$} (when task is complete). The Human uses \texttt{$action$} for structured commands and \texttt{$speak$} for natural language replies. Detailed prompts and procedures are in the Appendix \ref{app:balanced120-evaluation}.

\begin{table*}[t]
\caption{
\textbf{Performance Across Nine Capabilities.} Each of the 120 tasks is annotated with required capabilities and evaluated using separate state-verifiable checks. For each capability, we report the percentage of passed evaluation points across all applicable tasks. Models are ordered by overall task pass rate in descending order. The nine capabilities are Object Recognition, Spatial Grounding, 
OCR, Long-horizon Memory, Planning, Knowledge, Error Detection, Proactivity, and Communication.
}
\label{tab_nine_capbility}
\centering
\resizebox{0.95\textwidth}{!}{%
\begin{minipage}{475.5pt}
\begingroup

\definecolor{CapPositive}{HTML}{288F77} 
\definecolor{CapNegative}{HTML}{CA6878} 
\definecolor{CapInk}{HTML}{20252B}
\definecolor{CapMuted}{HTML}{5C636B}

\newcommand{\CapMaxTint}{74}
\newcommand{\CapRange}{18}

\newcommand{\CapRank}[2]{%
  \ifcase#2 #1\or\textbf{#1}\or\underline{#1}\fi
}

\newcommand{\CapHead}[2]{%
  \makebox[#1pt][c]{\textbf{#2}}%
}

\newcommand{\CapCell}[4]{%
  \begingroup
  \pgfmathsetmacro{\CapTint}{%
    min(1,abs(#3)/\CapRange)*\CapMaxTint%
  }%
  \ifdim #3pt<0pt
    \colorlet{CapFill}{CapNegative!\CapTint!white}%
  \else
    \colorlet{CapFill}{CapPositive!\CapTint!white}%
  \fi
  \setlength{\fboxsep}{0pt}%
  \colorbox{CapFill}{%
    \makebox[#1pt][c]{%
      \rule[-3.3pt]{0pt}{13.3pt}%
      \CapRank{#2}{#4}%
    }%
  }%
  \endgroup
}

\newcommand{\CapAverage}[2]{%
  \makebox[37pt][c]{\CapRank{#1}{#2}}%
}

\newcommand{\CapLegend}[1]{%
  \begin{tikzpicture}[x=1pt,y=1pt,baseline=0pt]
    \shade[
      left color=CapNegative!\CapMaxTint!white,
      right color=white
    ] (0,0) rectangle (64,5);

    \shade[
      left color=white,
      right color=CapPositive!\CapMaxTint!white
    ] (64,0) rectangle (128,5);

    \foreach \xx/\lab in #1 {
      \draw[CapMuted,line width=.3pt]
        (\xx,0)--(\xx,-2);
      \node[
        anchor=north,
        inner sep=0pt,
        text=CapMuted,
        font=\fontsize{6.7}{7}\selectfont
      ] at (\xx,-3.3) {\lab};
    }
  \end{tikzpicture}%
}

\fontsize{8.5}{10}\selectfont
\color{CapInk}
\setlength{\tabcolsep}{0pt}
\setlength{\aboverulesep}{3.2pt}
\setlength{\belowrulesep}{3.2pt}
\renewcommand{\arraystretch}{1}

\begin{tabular}{
  @{}l
  @{\hspace{4pt}}c
  @{\hspace{1.7pt}}c
  @{\hspace{1.7pt}}c
  @{\hspace{6pt}}c
  @{\hspace{6pt}}c
  @{\hspace{1.7pt}}c
  @{\hspace{1.7pt}}c
  @{\hspace{6pt}}c
  @{\hspace{1.7pt}}c
  @{\hspace{8pt}}c
  @{}
}

\toprule[.7pt]

& \multicolumn{3}{c}{\textbf{Perception}}
& \textbf{Memory}
& \multicolumn{3}{c}{\textbf{Reasoning}}
& \multicolumn{2}{c}{\textbf{Interaction}}
& \textbf{Overall} \\

\cmidrule(lr){2-4}
\cmidrule(lr){5-5}
\cmidrule(lr){6-8}
\cmidrule(lr){9-10}
\cmidrule(l){11-11}

\makebox[75pt][l]{\textbf{Model}}
& \CapHead{33}{Objects}
& \CapHead{33}{Spatial}
& \CapHead{33}{OCR}
& \CapHead{36}{Memory}
& \CapHead{33}{Plan}
& \CapHead{43}{Knowledge}
& \CapHead{34}{Detect}
& \CapHead{44}{Proactivity}
& \CapHead{34}{Comm.}
& \CapHead{37}{Avg.} \\

\midrule[.4pt]

Claude Fable 5.1
& \CapCell{33}{63.0}{8.13333333}{2}
& \CapCell{33}{61.2}{8.45833333}{2}
& \CapCell{33}{62.5}{6.04166667}{0}
& \CapCell{36}{58.9}{15.26666667}{1}
& \CapCell{33}{56.0}{9.81666667}{1}
& \CapCell{43}{54.9}{6.77500000}{0}
& \CapCell{34}{51.9}{7.10000000}{0}
& \CapCell{44}{44.4}{6.35833333}{0}
& \CapCell{34}{79.6}{5.86666667}{2}
& \CapAverage{59.2}{1} \\
\noalign{\vskip1.2pt}

Claude Opus 5
& \CapCell{33}{54.9}{0.03333333}{0}
& \CapCell{33}{57.1}{4.35833333}{0}
& \CapCell{33}{59.8}{3.34166667}{0}
& \CapCell{36}{46.4}{2.76666667}{0}
& \CapCell{33}{55.5}{9.31666667}{2}
& \CapCell{43}{54.4}{6.27500000}{0}
& \CapCell{34}{54.9}{10.10000000}{2}
& \CapCell{44}{53.4}{15.35833333}{1}
& \CapCell{34}{80.1}{6.36666667}{1}
& \CapAverage{57.4}{0} \\
\noalign{\vskip1.2pt}

Muse Spark 1.1
& \CapCell{33}{63.3}{8.43333333}{1}
& \CapCell{33}{62.0}{9.25833333}{1}
& \CapCell{33}{64.3}{7.84166667}{2}
& \CapCell{36}{48.9}{5.26666667}{0}
& \CapCell{33}{48.1}{1.91666667}{0}
& \CapCell{43}{54.2}{6.07500000}{0}
& \CapCell{34}{45.5}{0.70000000}{0}
& \CapCell{44}{33.5}{-4.54166667}{0}
& \CapCell{34}{76.5}{2.76666667}{0}
& \CapAverage{55.1}{0} \\
\noalign{\vskip1.2pt}

GPT-5.6 Sol
& \CapCell{33}{53.3}{-1.56666667}{0}
& \CapCell{33}{54.1}{1.35833333}{0}
& \CapCell{33}{48.7}{-7.75833333}{0}
& \CapCell{36}{49.2}{5.56666667}{0}
& \CapCell{33}{49.6}{3.41666667}{0}
& \CapCell{43}{43.2}{-4.92500000}{0}
& \CapCell{34}{51.1}{6.30000000}{0}
& \CapCell{44}{40.2}{2.15833333}{0}
& \CapCell{34}{78.5}{4.76666667}{0}
& \CapAverage{52.0}{0} \\
\noalign{\vskip1.2pt}

GPT-5.5
& \CapCell{33}{52.8}{-2.06666667}{0}
& \CapCell{33}{54.1}{1.35833333}{0}
& \CapCell{33}{47.7}{-8.75833333}{0}
& \CapCell{36}{40.2}{-3.43333333}{0}
& \CapCell{33}{44.7}{-1.48333333}{0}
& \CapCell{43}{45.7}{-2.42500000}{0}
& \CapCell{34}{46.6}{1.80000000}{0}
& \CapCell{44}{34.0}{-4.04166667}{0}
& \CapCell{34}{78.0}{4.26666667}{0}
& \CapAverage{49.3}{0} \\
\noalign{\vskip1.2pt}

GLM 5.3 Flash
& \CapCell{33}{59.3}{4.43333333}{0}
& \CapCell{33}{56.7}{3.95833333}{0}
& \CapCell{33}{69.3}{12.84166667}{1}
& \CapCell{36}{49.9}{6.26666667}{0}
& \CapCell{33}{54.6}{8.41666667}{0}
& \CapCell{43}{61.3}{13.17500000}{1}
& \CapCell{34}{56.8}{12.00000000}{1}
& \CapCell{44}{45.9}{7.85833333}{0}
& \CapCell{34}{69.8}{-3.93333333}{0}
& \CapAverage{58.2}{2} \\
\noalign{\vskip1.2pt}

Claude Opus 4.8
& \CapCell{33}{48.6}{-6.26666667}{0}
& \CapCell{33}{49.6}{-3.14166667}{0}
& \CapCell{33}{61.8}{5.34166667}{0}
& \CapCell{36}{33.9}{-9.73333333}{0}
& \CapCell{33}{43.7}{-2.48333333}{0}
& \CapCell{43}{40.7}{-7.42500000}{0}
& \CapCell{34}{36.4}{-8.40000000}{0}
& \CapCell{44}{48.5}{10.45833333}{2}
& \CapCell{34}{72.9}{-0.83333333}{0}
& \CapAverage{48.5}{0} \\
\noalign{\vskip1.2pt}

Kimi K3
& \CapCell{33}{59.1}{4.23333333}{0}
& \CapCell{33}{51.8}{-0.94166667}{0}
& \CapCell{33}{63.8}{7.34166667}{0}
& \CapCell{36}{50.1}{6.46666667}{2}
& \CapCell{33}{45.9}{-0.28333333}{0}
& \CapCell{43}{59.7}{11.57500000}{2}
& \CapCell{34}{53.9}{9.10000000}{0}
& \CapCell{44}{41.9}{3.85833333}{0}
& \CapCell{34}{74.8}{1.06666667}{0}
& \CapAverage{55.7}{0} \\
\noalign{\vskip1.2pt}

Qwen 3.8 Max
& \CapCell{33}{43.8}{-11.06666667}{0}
& \CapCell{33}{43.5}{-9.24166667}{0}
& \CapCell{33}{47.2}{-9.25833333}{0}
& \CapCell{36}{39.6}{-4.03333333}{0}
& \CapCell{33}{39.5}{-6.68333333}{0}
& \CapCell{43}{47.1}{-1.02500000}{0}
& \CapCell{34}{32.2}{-12.60000000}{0}
& \CapCell{44}{32.1}{-5.94166667}{0}
& \CapCell{34}{62.6}{-11.13333333}{0}
& \CapAverage{43.1}{0} \\
\noalign{\vskip1.2pt}

Gemini 3.1 Pro
& \CapCell{33}{55.3}{0.43333333}{0}
& \CapCell{33}{46.3}{-6.44166667}{0}
& \CapCell{33}{47.5}{-8.95833333}{0}
& \CapCell{36}{37.1}{-6.53333333}{0}
& \CapCell{33}{37.4}{-8.78333333}{0}
& \CapCell{43}{35.3}{-12.82500000}{0}
& \CapCell{34}{38.6}{-6.20000000}{0}
& \CapCell{44}{30.7}{-7.34166667}{0}
& \CapCell{34}{71.6}{-2.13333333}{0}
& \CapAverage{44.4}{0} \\
\noalign{\vskip1.2pt}

GPT-5.6 Luna
& \CapCell{33}{47.6}{-7.26666667}{0}
& \CapCell{33}{50.5}{-2.24166667}{0}
& \CapCell{33}{53.8}{-2.65833333}{0}
& \CapCell{36}{38.3}{-5.33333333}{0}
& \CapCell{33}{43.5}{-2.68333333}{0}
& \CapCell{43}{40.1}{-8.02500000}{0}
& \CapCell{34}{36.4}{-8.40000000}{0}
& \CapCell{44}{30.0}{-8.04166667}{0}
& \CapCell{34}{74.5}{0.76666667}{0}
& \CapAverage{46.1}{0} \\
\noalign{\vskip1.2pt}

DeepSeek V4.1 Flash
& \CapCell{33}{57.4}{2.53333333}{0}
& \CapCell{33}{46.0}{-6.74166667}{0}
& \CapCell{33}{51.1}{-5.35833333}{0}
& \CapCell{36}{31.1}{-12.53333333}{0}
& \CapCell{33}{35.7}{-10.48333333}{0}
& \CapCell{43}{40.9}{-7.22500000}{0}
& \CapCell{34}{33.3}{-11.50000000}{0}
& \CapCell{44}{21.9}{-16.14166667}{0}
& \CapCell{34}{65.9}{-7.83333333}{0}
& \CapAverage{42.6}{0} \\

\bottomrule[.7pt]
\end{tabular}

\par\vspace{6.2pt}

\begin{minipage}[t]{314pt}
  \fontsize{7.2}{9.2}\selectfont
  \color{CapMuted}
  Numbers: scores (\%).
  \textbf{Bold}: best.
  \underline{Underline}: second best.
  \par
  Green: above the capability mean\quad
  Red: below the capability mean\quad
  All 12 models.
\end{minipage}\hfill
\begin{minipage}[t]{168pt}
  \centering
  \fontsize{7.0}{9.2}\selectfont
  \color{CapMuted}
  Deviation from capability mean (pp)
  \par\vspace{3pt}
  \CapLegend{{
    0/$-18$,
    21.333/$-12$,
    42.667/$-6$,
    64/$0$,
    85.333/$+6$,
    106.667/$+12$,
    128/$+18$
  }}
\end{minipage}

\endgroup
\end{minipage}}
\end{table*}

\subsection{Benchmark Results and Analysis}
\label{sec:main_results}
 
\begin{figure}
 \centering
 \centerline{\includegraphics[width=0.95\linewidth]{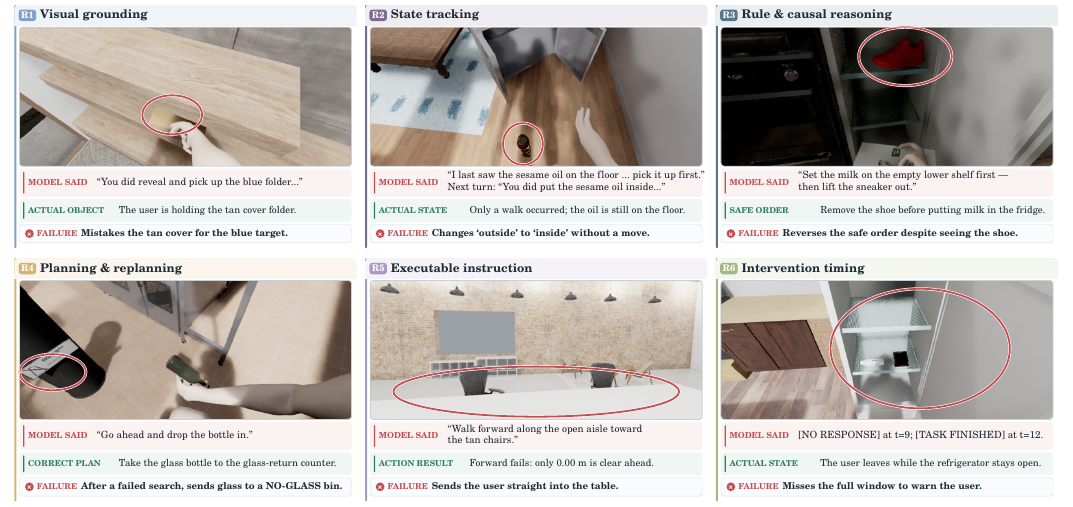}}
 \vspace{-3pt}
 \centerline{\includegraphics[width=0.95\linewidth]{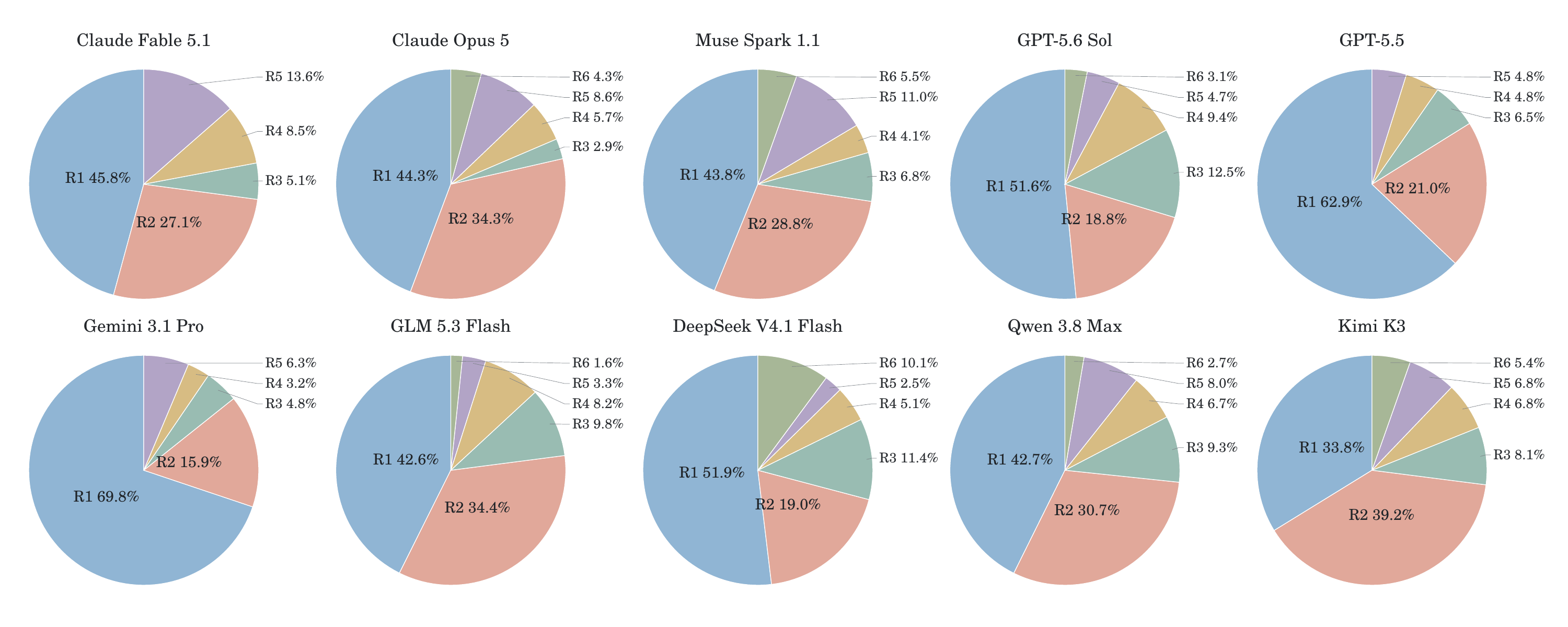}}
 \vspace{-10pt}
 \caption{\textbf{Examples and Distribution of Error Types.} We analyze initial failure of failed tasks and grouped into six  types. (Top) Examples of each error type. (Bottom) Error type distributions.}
 \label{fig_error}
\end{figure}

We evaluate models on InteractiveBench through five lenses:
(i)~\textbf{Performance}: measuring the Human's task success rate and mean completion rate (Tab.~\ref{table_pass_rate});
(ii)~\textbf{Efficiency}: calculating the ratio of actual human actions to those required with an oracle assistant (Fig.~\ref{fig_efficiency});
(iii)~\textbf{Capability Assessment}: evaluating nine capabilities (perception, memory, reasoning, and interaction) based on predefined state conditions (Tab.~\ref{tab_nine_capbility});
(iv)~\textbf{Failure Analysis}: classifying the initial error in failed tasks into six categories (Fig.~\ref{fig_error}); and
(v)~\textbf{Complexity Impact}: analyzing the correlation between success rates and task length (Fig.~\ref{fig_horizon}).
From these evaluations (details in Appendix \ref{app:experiment}), we distill five key findings.

\finding{find:overall}{Closed-loop assistance is far from solved: even current strongest MLLM cannot yet serve as a reliable egocentric AI Assistant, helping the Human complete fewer than half of the tasks.}

As shown in Tab.~\ref{table_pass_rate}, Claude Fable 5.1 achieves the highest average success rate at only 41.7\%, with all other models at or below 33.3\%. Since we verified that an Oracle Assistant achieves a 100\% success rate with the same Human agent, this substantial gap indicates that current models still fail to provide reliable guidance, leaving immense room for improvement in collaborative AI.

\finding{find:loop}{The difficulty lies in sustaining the loop: models make good one-shot decisions but break down when they must keep tracking progress and re-planning.}
 
Performance varies significantly by task category. While best models reach an 85.0\% success rate on Decision tasks, they struggle heavily on Guidance (20.0\%) and Planning (25.0\%). This gap stems from task complexity: Decision tasks often rely on a single scene judgment, whereas Guidance and Planning demand continuous tracking, anomaly detection, and dynamic replanning. Consistent with this, Fig.~\ref{fig_horizon} shows that model performance drops quickly as tasks get longer, especially on guidance and planning. Together, these results show what existing static evaluations fail to capture.
 
\finding{find:perception_memory}{Most failures stem from perception and memory: models misread scenes or lose track of changes, revealing current models' poor foundational visual capabilities for interaction.}
 
Across all models, visual perception and memory errors account for 70.4\%--85.7\% of first errors (Fig.~\ref{fig_error}). Because closed-loop interaction requires continuously tracking dynamic scenes, a single misperception or forgotten state cascades into all subsequent instructions. This result indicates that current models' foundational capabilities for interaction remain insufficient. Furthermore, the breakdown also varies across models: visual perception accounts for 69.8\% of Gemini 3.1 Pro's first errors, whereas Kimi K3 fails more often on memory than on perception (39.2\% vs.\ 33.8\%).

\finding{find:guidance}{Communication is critical: if the Assistant cannot express its plan as clear, step-by-step guidance, strong performance in other capabilities does little to help the Human.}
 
Models with higher success rates generally score higher on most capabilities (Tab.~\ref{tab_nine_capbility}), but the average does not determine success. GLM-5.3-Flash and Claude Fable 5.1 have nearly identical average capability scores (58.2\% vs.\ 59.2\%), yet GLM-5.3-Flash scores markedly lower on communication (69.8\% vs.\ 79.6\%) and completes 11.7 points fewer tasks (30.0\% vs.\ 41.7\%). The error analysis points the same way: even for Claude Fable 5.1, 13.6\% of first errors arise when the Assistant fails to turn the task goal into guidance the Human can follow (R5 example in Fig.~\ref{fig_error}). Closed-loop success is thus bounded by the weakest capability in the chain rather than by the average, and communication of this kind is not measured by static question-answering benchmarks before.
 
\finding{find:efficiency}{Action efficiency also remains low: current models lead the Human through many unnecessary actions, making efficiency an important dimension for improving assistance alongside task success.}
 
Compared to the Oracle, almost all models exhibit severe action redundancy (Fig.~\ref{fig_efficiency}), peaking in Accessibility and Guidance (where models require $2.35\times$--$3.66\times$ the Oracle's cost). Furthermore, efficiency is highly task-dependent rather than a fixed model trait. For instance, Claude Fable 5.1, despite being the most successful overall, is the least efficient on Accessibility ($6.75\times$), yet highly efficient on Prevention ($1.20\times$). Similarly, Kimi K3 matches the Oracle on Finding ($1.00\times$) but suffers the highest cost on Guidance ($3.66\times$). Since unnecessary instructions directly translate to wasted human effort, optimizing interaction efficiency is as critical as maximizing success rates.
 
\textbf{Emergent proactive behaviors.} Beyond task success and efficiency, we also observe that models exhibit proactive information-seeking behaviors that are not explicitly required by the Assistant prompt. First, when the surrounding objects are unclear, models ask the Human to walk around or look around the space to obtain a clearer view of the scene. Second, they ask the Human to flip packages or other objects to reveal labels and information that is otherwise hidden. Third, they actively ask about spatial information, such as where objects are located relative to the Human, to better ground their guidance. Instead of passively waiting for informative frames, these models actively shape what they observe through the Human. Such behaviors are an essential part of helping people in the physical world, where key information is often occluded or outside the current view.

\subsection{Ablation Study}
Furthermore, we conduct ablation studies to investigate (1) the effectiveness of inference-time memory augmentation, and (2) the impact of varying the Human agent model on task performance.

\textbf{Inference-time memory augmentation.} To address state-tracking limitations, we append a hidden \texttt{\$Memory\$} field to the Assistant's responses, allowing it to record events for subsequent turns. This augmentation improves success rates by 5--15\% on Prevention and Finding across all three models (Fig.~5). However, its effects on Decision, Guidance, and Accessibility vary across models, while Planning success drops by 5\% for two models. These mixed results suggest that this simple memory augmentation alone is insufficient for reliable long-horizon assistance, motivating future work on training native memory mechanisms for dynamic interaction. See Appendix~\ref{app:experiment} for details.

\begin{figure*}[t]
\centering

\begin{minipage}{0.28\textwidth}

    \centering
    \includegraphics[width=\linewidth]{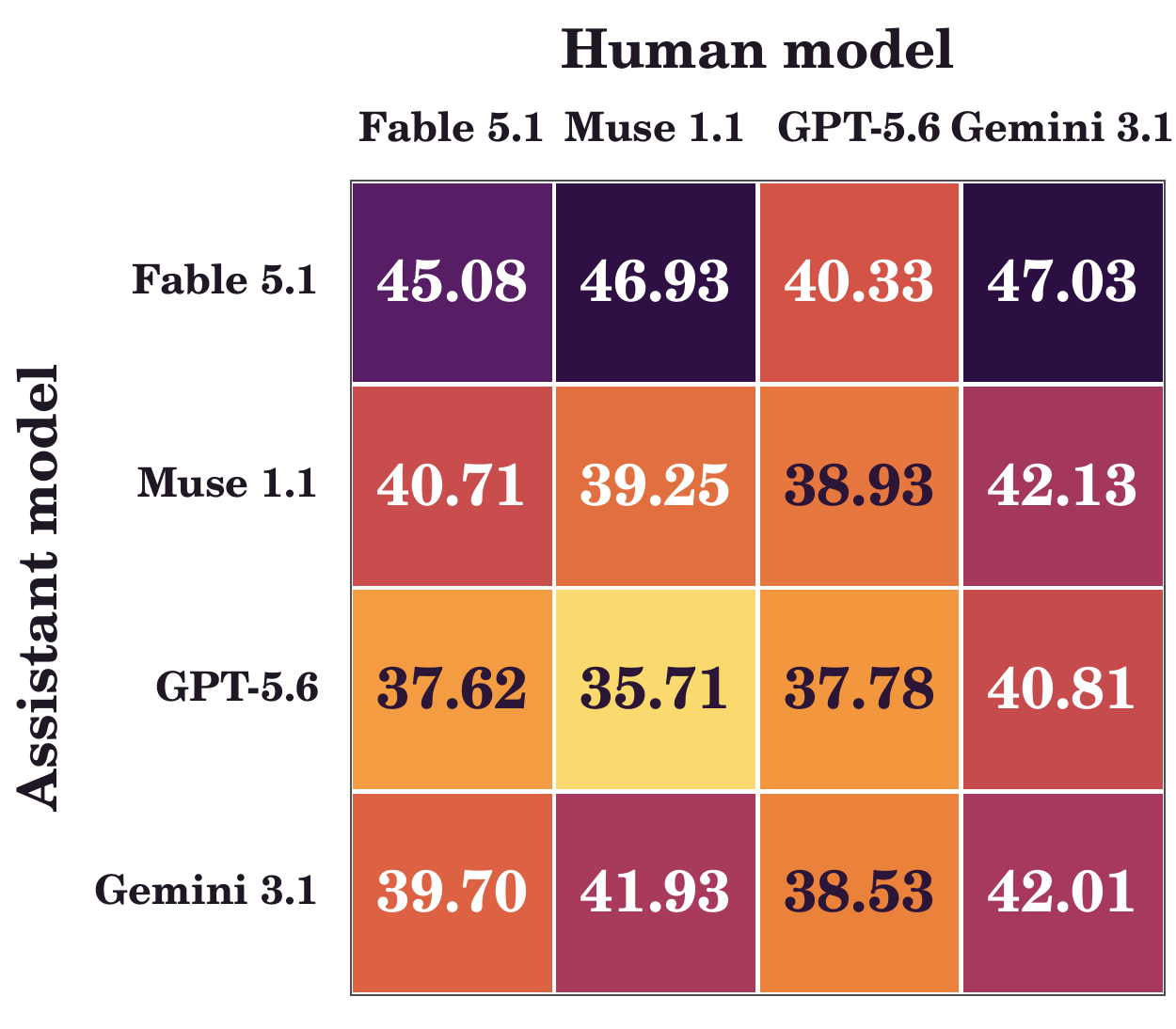}
\end{minipage}
\hspace{3pt}
\begin{minipage}{0.65\textwidth}
    \centering
    \small
    \resizebox{\linewidth}{!}{%
    \begin{tabular}{lcccccc}
        \toprule[1.2pt]
        \textbf{Model} & \textbf{Guide} & \textbf{Plan} & \textbf{Access.}
        & \textbf{Decide}& \textbf{Preven.} & \textbf{Find} \\
        \midrule
        \midrule
        Fable 5.1
        & 20.0\% & 25.0\%  & 30.0\%  & 85.0\%  & 50.0\%  & 40.0\%   \\
        \rowcolor{memorygray}
        + Memory
        & 15.0\%  & 20.0\%  & 30.0\%  & 90.0\%  & 55.0\%  & 50.0\%  \\
        \midrule
        GPT-5.6-sol
        & 20.0\%  & 25.0\%  & 30.0\%  & 70.0\%  & 15.0\%  & 25.0\%  \\
        \rowcolor{memorygray}
        + Memory
        & 20.0\%  & 20.0\%  & 15.0\%  & 65.0\%  & 30.0\%  & 40.0\%  \\
        \midrule
        Gemini 3.1 Pro
        & 10.0\%  & 10.0\%  & 35.0\%  & 50.0\%  & 10.0\%  & 10.0\%  \\
        \rowcolor{memorygray}
        + Memory
        & 15.0\%  & 10.0\%  & 45.0\%  & 55.0\%  & 15.0\%  & 20.0\%  \\
        \bottomrule[1.2pt]
    \end{tabular}%
    }
    
\end{minipage}
\label{tab:memory_ablation}
\caption{
(Left) \textbf{Human Agent Impact.} We report average MCR across Guidance, Planning, and Accessibility (tasks requiring maximal human involvement) when cross-evaluating four diverse models interchangeably as assistant and human agents. (Right) \textbf{Memory Augmentation Impact.} We report model pass rates with vs. without inference-time memory augmentation.
}

\end{figure*}

\begin{figure*}
 \centering
 \centerline{\includegraphics[width=0.96\linewidth]{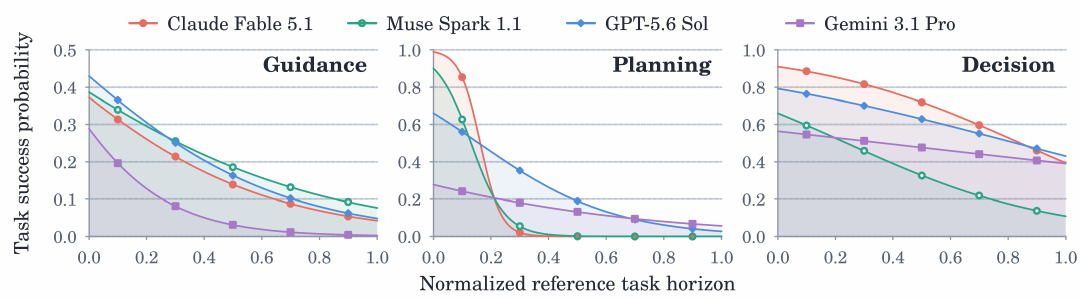}}
 \caption{\textbf{Performance Across Task Horizons.} We report model success rates against task lengths (derived from oracle actions). Task horizons are normalized per category on the x-axis (0: shortest, 1: longest), while the y-axis shows the average pass rate at each normalized length step.}
 \label{fig_horizon}
\end{figure*}

\textbf{Effect of the Human agent.} To evaluate the Human's impact, we test all 16 Assistant--Human pairings among four diverse models, reporting MCR on Guidance, Planning, and Accessibility, long-horizon tasks reliant on instruction following (Fig.~5). An Assistant's performance varies with its Human partner, although rankings remain relatively stable: Fable 5.1 consistently ranks first, while GPT-5.6 ranks last. Performance also differs across roles: although Gemini 3.1 ranks second or third as an Assistant, all four Assistants achieve their highest observed MCR when it acts as the Human. This implies that the challenge in Human agent is different from the Assistant agent and Gemini can be a good proxy of Human. Future work will explore behavioral calibration to better align simulated Human agents with real users.


\section{Conclusion}
We introduced InteractiveWorld, a verifiable environment for closed-loop human-assistant interaction, and InteractiveBench, a dynamic benchmark of 120 multi-turn tasks. InteractiveWorld measures interactive abilities with verifiable rewards, while InteractiveBench evaluates how models assist people in real world. They provide a foundation for developing and evaluating interactive MLLMs.

\bibliographystyle{assets/plainnat}
\bibliography{references}

\newpage
\beginappendix

\newcommand{\TOCentry}[3][0em]{%
  \noindent\hspace*{#1}%
  \hyperref[#2]{#3}%
  \TOCdots
  \hyperref[#2]{\pageref{#2}}%
  \par
}

\newcommand{\TOClineAuto}[2]{\TOCentry[0em]{#2}{#1}}
\newcommand{\TOCsublineAuto}[2]{\TOCentry[1.5em]{#2}{#1}}

\section*{Appendix Contents}
\TOClineAuto{A Background \& Related Work}{app:background}

\TOClineAuto{B Implementation Details of InteractiveWorld}{app:interactiveworld}
\TOCsublineAuto{B.1 Implementation Details of Physical Layer }{app:simulator-physical}
\TOCsublineAuto{B.2 Implementation Details of Human Action Layer }{app:simulator-human-action}
\TOCsublineAuto{B.3 Implementation Details of Semantic Layer }{app:simulator-semantic}

\TOClineAuto{C Implementation Details of InteractiveBench}{app:interactivebench}
\TOCsublineAuto{C.1 Benchmark Construction}{app:balanced120-construction}
\TOCsublineAuto{C.2 Evaluation Protocol}{app:balanced120-evaluation}
\TOCsublineAuto{C.3 Evaluation Metrics}{app:balanced120-metrics}
\TOCsublineAuto{C.4 Task Catalog}{app:balanced120-tasks}

\TOClineAuto{D Evaluation Setting Details and More Experiment Results}{app_error}
\TOCsublineAuto{D.1 Experiment Setting Detail of Error Type Analysis}{app_error}
\TOCsublineAuto{D.2 Experiment Setting Detail of Memory Augmentation}{app_memory}
\TOCsublineAuto{D.3 Additional Experiments about Comparative of Dynamic vs. Static}{app:static-benchmark}
\TOCsublineAuto{D.4 Additional Experiment of Task Horizon and Model Performance}{sec:horizon_analysis}
\TOCsublineAuto{D.5 Additional Human-Agent Validation with Real Human Participants}{app:human_validation}

\TOClineAuto{E Future Work}{app:futurework}



\section{Background \& Related Work}
\label{app:background}

\noindent\textbf{Egocentric AI Assistant Benchmarks.}
Large-scale first-person datasets such as EPIC-KITCHENS~\cite{damen2018epic}, Ego4D~\cite{grauman2022ego4d}, and Ego-Exo4D~\cite{grauman2024egoexo4d} have laid the foundation for egocentric perception.
Building on them, a growing body of benchmarks examines whether MLLMs can act as egocentric assistants: EgoSchema~\cite{angalam2023egoschema} and HourVideo~\cite{chandrasegaran2024hourvideo} test long-form video understanding; EgoThink~\cite{cheng2024egothink} and VidEgoThink~\cite{cheng2024videgothink} evaluate first-person reasoning; EgoPlan-Bench~\cite{chen2023egoplan} targets next-step planning; EgoLife~\cite{yang2025egolife} and TeleEgo~\cite{yan2025teleego} probe long-term memory over continuously recorded daily life; and EgoBlind~\cite{xiao2025egoblind} focuses on visual assistance for blind users.
Closer to task guidance, HoloAssist~\cite{wang2023holoassist} records real-world instructor--performer collaborations, while ProAssist~\cite{zhang2025proassist} and the ESTP task~\cite{zhang2025eyeswideopen} evaluate when and what a proactive assistant should say along a streaming egocentric video.
Despite their diversity, these benchmarks are \emph{open-loop}: model outputs are scored against fixed recordings and never influence what happens next.
EgoBench~\cite{liu2026egobench} introduces an LLM-simulated user and tool calls, yet its visual context remains pre-recorded.
User-in-the-loop studies~\cite{verghese2025userintheloop,dasgupta2026puzzlemate} do close the loop with real participants, and Verghese et al.~\cite{verghese2025userintheloop} further show that offline metrics do not reflect online performance; however, such studies involve only a handful of participants and are costly to scale and hard to reproduce.
InteractiveBench closes this loop in simulation: every instruction changes the human's actions and thus the world, and success is verified automatically from final object states.

\noindent\textbf{Improving Egocentric AI Assistants.}
In parallel, many works improve the models and frameworks behind egocentric assistants.
Early efforts learn egocentric video--language representations from narrated first-person videos~\cite{lin2022egovlp,zhao2023lavila}, and subsequent MLLMs such as MM-Ego~\cite{ye2025mmego}, AlanaVLM~\cite{suglia2024alanavlm}, and EgoGPT~\cite{yang2025egolife} are tuned on large-scale egocentric instruction data.
To handle continuous input, VideoLLM-online~\cite{chen2024videollmonline}, Dispider~\cite{qian2025dispide}, and proactive assistant models~\cite{zhang2025proassist,zhang2025eyeswideopen} learn to decide when to respond as the video unfolds.
For long-horizon memory, LifelongMemory~\cite{wang2023lifelongmemory} and EgoRAG~\cite{yang2025egolife} retrieve from textualized histories, and Ego-R1~\cite{tian2025egor1} trains an agent with reinforcement learning to invoke retrieval and perception tools step by step over week-long videos.
At the system level, VLaMP~\cite{patel2023vlamp} leverages pretrained language models to plan the next steps for human assistance, while Vinci~\cite{huang2024vinci} and Egocentric Co-Pilot~\cite{yang2026egocopilot} integrate memory, planning, and tool use into assistants deployable on portable devices.
However, these methods are trained and evaluated on static data or small-scale user studies, so it remains unclear whether their gains carry over to closed-loop assistance, where the assistant must track how its own guidance changes the scene and the human.
InteractiveWorld offers a verifiable testbed to answer this question, and its state-based rewards can further support training interactive assistants.

\definecolor{iwGreen}{RGB}{16,150,72}
\definecolor{iwRed}{RGB}{214,39,40}
\definecolor{iwOrange}{RGB}{240,140,0}
\definecolor{iwOurs}{RGB}{232,237,252}
\definecolor{iwGroup}{RGB}{90,90,90}
\DeclareRobustCommand{\iwY}{\textcolor{iwGreen}{\ding{51}}}   
\DeclareRobustCommand{\iwN}{\textcolor{iwRed}{\ding{55}}}     
\DeclareRobustCommand{\iwP}{\tikz[baseline=-0.68ex]{
  \draw[iwOrange,line width=0.6pt] (0,0) circle (0.72ex);%
  \fill[iwOrange] (0,0.72ex) arc (90:270:0.72ex) -- cycle;}}
\newcommand{\iwGroupRow}[1]{\multicolumn{10}{l}{\textcolor{iwGroup}{\textit{#1}}}\\}
 
\begin{table*}[t]
\centering
\caption{\textbf{Comparison with existing egocentric AI assistant benchmarks and embodied simulators.}
InteractiveWorld is the only environment that satisfies all six properties.
\emph{Ego.\ View}: first-person observations of a human.
\emph{Human Actor}: the task is physically performed by a (real or simulated) human, not a robot.
\emph{Guided Human}: the human autonomously grounds the agent's free-form language guidance into actions.
\emph{NL Dialog}: multi-turn natural-language interaction.
\emph{Closed Loop}: agent outputs change the environment and the subsequent observations.
\emph{Verif.\ Success}: success is checked against ground-truth environment states rather than by answer matching or LLM judges.
\iwP: partially supported.}
\label{tab:comparison}
\small
\setlength{\tabcolsep}{3.5pt}
\renewcommand{\arraystretch}{1.12}
\begin{adjustbox}{max width=\textwidth}
\begin{tabular}{llll cccccc}
\toprule
\multirow{2}{*}{\textbf{Method}} &
\multirow{2}{*}{\textbf{Primary Focus}} &
\multirow{2}{*}{\textbf{Platform}} &
\multirow{2}{*}{\textbf{Agent Output}} &
\multicolumn{3}{c}{\textbf{Human-Centric}} &
\multicolumn{3}{c}{\textbf{Interaction \& Eval.}} \\
\cmidrule(lr){5-7}\cmidrule(l){8-10}
 & & & &
\shortstack{Ego.\\View} & \shortstack{Human\\Actor} & \shortstack{Guided\\Human} &
\shortstack{NL\\Dialog} & \shortstack{Closed\\Loop} & \shortstack{Verif.\\Success} \\
\midrule
\iwGroupRow{Egocentric AI assistant benchmarks (offline, pre-recorded real-world videos)}
EgoSchema~\cite{angalam2023egoschema}   & Long-form video QA        & Ego4D          & MCQ answer        & \iwY & \iwY & \iwN & \iwN & \iwN & \iwN \\
EgoThink~\cite{cheng2024egothink}        & First-person VQA          & Ego4D          & Free-form answer  & \iwY & \iwY & \iwN & \iwN & \iwN & \iwN \\
EgoPlan-Bench~\cite{chen2023egoplan}     & Egocentric planning       & Ego4D, EPIC    & MCQ (next step)   & \iwY & \iwY & \iwN & \iwN & \iwN & \iwN \\
EgoLife~\cite{yang2025egolife}           & Long-term life QA         & Aria glasses   & MCQ answer        & \iwY & \iwY & \iwN & \iwN & \iwN & \iwN \\
HoloAssist~\cite{wang2023holoassist}     & Task guidance             & HoloLens 2     & Labels / poses    & \iwY & \iwY & \iwP & \iwP & \iwN & \iwN \\
ProAssist~\cite{zhang2025proassist}      & Proactive guidance        & 6 datasets     & Dialogue turns    & \iwY & \iwY & \iwN & \iwY & \iwN & \iwN \\
\midrule
\iwGroupRow{Robot-centric embodied simulators}
Habitat 2.0~\cite{szot2021habitat}       & Rearrangement             & Habitat        & Robot actions     & \iwN & \iwN & \iwN & \iwN & \iwY & \iwY \\
BEHAVIOR-1K~\cite{li2023behavior}        & Household activities      & OmniGibson     & Robot actions     & \iwN & \iwN & \iwN & \iwN & \iwY & \iwY \\
ManiSkill3~\cite{tao2025maniskill3}      & Manipulation              & SAPIEN         & Robot actions     & \iwN & \iwN & \iwN & \iwN & \iwY & \iwY \\
Isaac Lab~\cite{mittal2025isaaclab}      & Robot learning            & Isaac Sim      & Robot actions     & \iwN & \iwN & \iwN & \iwN & \iwY & \iwY \\
\midrule
\iwGroupRow{Human- and interaction-oriented simulators}
VirtualHome~\cite{puig2018virtualhome}   & Activity programs         & Unity          & Action scripts    & \iwY & \iwY & \iwN & \iwN & \iwY & \iwY \\
TEACh~\cite{padmakumar2022teach}         & Dialogue-guided tasks     & AI2-THOR       & Follower actions  & \iwP & \iwN & \iwP & \iwY & \iwY & \iwY \\
Habitat 3.0~\cite{puig2024habitat3}      & Human--robot collab.      & Habitat        & Robot actions     & \iwN & \iwP & \iwN & \iwN & \iwY & \iwY \\
GRUtopia~\cite{wang2024grutopia}         & City-scale society        & Isaac Sim      & Robot actions     & \iwN & \iwN & \iwN & \iwY & \iwY & \iwY \\
FreeAskWorld~\cite{peng2026freeaskworld} & Direction inquiry         & Unity          & Nav.\ + queries   & \iwN & \iwN & \iwN & \iwY & \iwY & \iwY \\
HumanCLAW~\cite{siyao2026humanclaw}      & Acting via a body         & Habitat        & Body skills       & \iwY & \iwY & \iwN & \iwN & \iwY & \iwY \\
\midrule
\rowcolor{iwOurs}
\textbf{InteractiveWorld (Ours)} & \textbf{Human--assistant collab.} & \textbf{Isaac Sim} & \textbf{NL guidance} & \iwY & \iwY & \iwY & \iwY & \iwY & \iwY \\
\bottomrule
\end{tabular}
\end{adjustbox}
\end{table*}

\noindent\textbf{Robot-Centric Embodied Simulators.}
Simulators have long provided scalable and reproducible testbeds for embodied AI.
AI2-THOR~\cite{kolve2017ai2thor}, Habitat~\cite{savva2019habitat,szot2021habitat}, iGibson~\cite{li2021igibson2}, and BEHAVIOR-1K~\cite{li2023behavior} offer interactive household scenes for navigation, rearrangement, and everyday activities; ManiSkill3~\cite{tao2025maniskill3} and Isaac Lab~\cite{mittal2025isaaclab} provide GPU-parallel physics and rendering for large-scale robot learning; and RoboCasa~\cite{nasiriany2024robocasa} and GRUtopia~\cite{wang2024grutopia} scale up scene diversity from kitchens to entire cities.
On top of these platforms, benchmarks such as ALFRED~\cite{shridhar2020alfred} and EmbodiedBench~\cite{yang2025embodiedbench} evaluate language-conditioned agents and MLLMs by having them directly command a robot.
In these settings, the evaluated model is the \emph{actor}: it outputs actions that the simulator executes on a robot body, while humans are either absent or merely part of the scene.
HumanCLAW~\cite{siyao2026humanclaw} replaces the robot with a physically simulated human body and decouples decision making from motor control, yet the VLM still decides every move of that body.
In contrast, the assistant in InteractiveWorld never acts on the world directly; it can influence the outcome only through natural-language guidance that an autonomous human must interpret and carry out, mirroring how assistants help people in practice.

\noindent\textbf{Interaction-Oriented Simulators.}
Another line of work brings humans or dialogue into simulation.
In dialog-based embodied tasks such as CVDN~\cite{thomason2020cvdn}, TEACh~\cite{padmakumar2022teach}, and DialFRED~\cite{gao2022dialfred}, an instruction giver guides a follower through natural language; however, models are typically evaluated as the follower, with guidance replayed from recorded human dialogues or produced by a rule-based oracle.
Human--AI collaboration platforms instead simulate human-like agents alongside AI helpers: VirtualHome~\cite{puig2018virtualhome} and Watch-And-Help~\cite{puig2021watchandhelp} model household activities of humanoid avatars; Habitat 3.0~\cite{puig2024habitat3} and PARTNR~\cite{chang2025partnr} study human--robot coordination with simulated humanoids and real humans in the loop; and CoELA~\cite{zhang2024coela}, Virtual Community~\cite{zhou2026virtualcommunity}, and HRIBench~\cite{liu2026hribench} extend this to LLM-driven cooperation, open-world societies, and role-conditioned interaction.
FreeAskWorld~\cite{peng2026freeaskworld} further employs LLM-driven avatars that give directions to a navigating agent.
Across these platforms, the AI remains the physical actor, while humans serve as scripted partners, co-workers, or sources of information.
InteractiveWorld reverses these roles: the AI is a disembodied assistant that shares the human's view, and an autonomous MLLM-driven human decides how to act on its guidance, enabling closed-loop, verifiable evaluation of assistants that help people rather than act for them.

\section{Implementation Details of InteractiveWorld}
\label{app:interactiveworld}

As mentioned in the main text, the simulator comprises three layers: a \textit{physical layer} that maintains a
geometrically consistent human, scene, and set of rigid objects
(\S\ref{app:simulator-physical}); a \textit{human-action layer} that converts
semantic goals into locomotion and manipulation trajectories
(\S\ref{app:simulator-human-action}); and a \textit{semantic layer} that exposes
these capabilities through a compact command language and a verifiable
multi-turn interaction loop (\S\ref{app:simulator-semantic}). 

\subsection{Implementation Details of Physical Layer}
\label{app:simulator-physical}

\subsubsection{Stable Human Representation}
The camera wearer is a fixed-shape SMPL+H model: the male model with zero
shape coefficients ($\boldsymbol{\beta}=\mathbf{0}$), a 52-joint kinematic
tree, 6,890 vertices, and 13,776 triangles.  Each frame is parameterized by a
156-D axis--angle vector
\begin{equation}
  \mathbf{q}
  = [\boldsymbol{\omega}_{\mathrm{root}},
     \boldsymbol{\theta}_{\mathrm{body}},
     \boldsymbol{\theta}_{\mathrm{left}},
     \boldsymbol{\theta}_{\mathrm{right}}]
  \in \mathbb{R}^{3+63+45+45},
\end{equation}
comprising the root orientation, 21 body joints (torso, head, arms, legs), and
15 articulated joints per hand; root translation is stored separately in world
coordinates.  A fixed $90^{\circ}$ rotation about $x$ maps the $y$-up model to
the $z$-up world.  Standard SMPL+H forward kinematics and linear blend skinning
yield joint transforms and deformed vertices.  The body mesh is authored once
at \texttt{/World/Body}; each frame updates only vertex positions and normals,
so topology, material binding, and camera attachment remain fixed.

Locomotion reuses body motion from ACCAD/AMASS SMPL-X clips: shared body
rotations are copied, incompatible shape parameters are discarded, and the hand
channels either keep a task-specific pose or are overwritten by the
manipulation controller.  The importer detects the source up axis,
canonicalizes the initial heading, and estimates the gait period by
autocorrelation.  Gait phase advances with distance traveled rather than
wall-clock time, so slowing, turning in place, or pausing to replan causes no
foot sliding; the lowest foot is continuously aligned with the estimated
support floor.  Manipulation poses are layered onto the same skeleton, avoiding
a model switch between walking, reaching, and object interaction.

The human is deliberately kinematic rather than a ragdoll, which makes the
camera trajectory reproducible and prevents falling or numerical drift under
contact.  Collisions are nevertheless enforced: every desired pose is screened
against an collision representation before it is committed.

\subsubsection{Collision-Free Interaction}

\paragraph{Proxy geometry.}
Rendering uses the full SMPL+H surface, whereas collision queries use a compact
proxy rig of 17 capsules (pelvis, spine, neck/collars, upper and lower arms,
hands, thighs, shins, feet) plus a head sphere.  Capsule endpoints are
regenerated from the current joints and radii are derived from the
corresponding body segments; hand capsules extend from the wrist toward the
palm.  Contacts between adjacent segments are filtered, while contacts with the
scene, manipulated objects, and non-adjacent body parts remain active.

\paragraph{Shared collision world.}
A single collision world serves scene initialization, navigation, stance
selection, reaching, and post-action validation.  It enumerates
collision-enabled scene geometry, follows USD instance proxies, preserves each
mesh's vertical extent, and groups multi-mesh articulated links under one
owning object.  Proposed poses are checked without one-frame lag by sampling
spheres along each capsule at the \textit{requested} pose rather than the last
rendered one.  Candidate stances are tested both in the idle configuration and
with the relevant arm extended; the intended contact region around the target
(nominal radius $0.22\,\mathrm{m}$) may be excluded from the reach test, but
the rest of the target and the scene remain obstacles.  On collision, bisection
estimates the penetration and the controller tries another stance, posture, or
approach.  Manipulation trajectories use a stricter acceptance tolerance of
${\approx}2\,\mathrm{mm}$.

\paragraph{Carried objects.}
The held object's collision geometry is propagated to every sampled wrist pose
and tested against the scene and all non-held objects; only the intended
hand--object grasp contact and the commanded target contact are allow-listed.
An arm-safe trajectory is therefore still rejected if the carried object would
strike a shelf, container rim, or neighboring object.  Placement candidates are
checked against non-target collision owners before release, and a post-settle
overlap check catches contacts introduced by rigid-body dynamics.

\paragraph{Object physics.}
Static architectural and furniture assets use their authored collision meshes
when available.  Movable BEHAVIOR objects use the asset's authored convex
pieces or, failing that, a convex decomposition of the concave visual mesh
rather than a single concave dynamic collider.  Initial packing enforces a
nominal $2\,\mathrm{cm}$ separation; objects are then settled under gravity
($9.81\,\mathrm{m\,s^{-2}}$, 60 physics steps/s) until their speed falls below
$10^{-3}\,\mathrm{m\,s^{-1}}$.  A grasped object becomes kinematic and follows
the wrist through a fixed relative transform.  On release it becomes dynamic,
is settled and, if necessary, minimally depenetrated, and its resulting
transform is written back to the semantic state.  Contents of a carried
container follow via their stored relative transforms.

\paragraph{Scope of the guarantee.}
``Collision-free'' denotes an operational guarantee: every planned route and
manipulation is screened against the same geometry and rejected or repaired if
it exceeds the configured contact tolerance.  It does not claim exact
separation between continuous triangle surfaces.  In the egocentric evaluation
configuration, small body-only contacts may be retained as diagnostic
telemetry provided the camera/eye volume stays clear, whereas an object
penetration above $5\,\mathrm{cm}$ fails the final placement gate.  This
prevents harmless skin-level numerical contacts from destabilizing behavior
while still rejecting visually or physically consequential interpenetration.

\subsubsection{Rich and Scalable Asset Support}
We use BEHAVIOR-1K v3.9.0, which contains 1,829 object categories, 8,635 model
directories, and 51 scene layouts.  The layouts place 13,516 object instances
covering 3,707 distinct category/model pairs.  The generic importer omits
purely decorative or non-interactive classes (e.g., ceilings, pictures,
mirrors, windows, curtains, lighting fixtures), leaving 11,428 instances and
3,189 distinct models, all present in the deployed cache.  At the time of this
audit, the cache holds 6,831 plain-USD models spanning all 1,829 categories.
Of the 51 layouts, 44 pass generic task admission (successful bake, usable
floors and walls, and at least one work surface or container).  BEHAVIOR's 996
material/particle systems are catalogued separately and are not counted as
manipulable rigid objects in our action interface.

The layouts span domestic, office, retail, educational, hospitality, and
leisure environments, e.g., \texttt{Beechwood\_0\_int}, \texttt{Ihlen\_0\_int},
\texttt{house\_single\_floor}, \texttt{office\_cubicles\_right},
\texttt{restaurant\_diner}, \texttt{school\_chemistry},
\texttt{grocery\_store\_cafe}, and \texttt{hotel\_gym\_spa}.  Manipulable
objects include apples, soda cans, bananas, bars of soap, tennis balls, bell
peppers, mugs, bowls, cabinet doors, and refrigerator contents.

Asset preparation is fully automated.  An offline baker decrypts the
distributed USD when necessary, re-roots PBR texture paths, removes
dataset-specific physics scenes and helper rigid bodies, and extracts
articulated-joint metadata.  It emits a plain USD and a compact specification
(category/model identifiers, bounding box, base offset, canonical orientation,
mass or density, openable links, semantic tags, and joints).  At load time, the
scene importer references this USD, applies the scene-JSON translation, scale,
and quaternion (converted from $xyzw$ to $wxyz$), aligns the asset base with
the floor, registers its semantic roles, and installs its collision geometry.
Adding a BEHAVIOR asset thus requires only baking and metadata validation, with
no simulator-code change.

\subsection{Implementation Details of Human Action Layer}
\label{app:simulator-human-action}

\subsubsection{Collision-Aware Autonomous Navigation}

\paragraph{Stance selection.}
Navigation targets an \textit{interaction-ready stance} rather than the object's
centroid.  Target grounding estimates a task-relevant contact point and samples
floor-supported standing poses around it, each oriented toward that point and
evaluated for (i)~connectivity to the current pose, (ii)~camera and full-body
clearance, (iii)~arm reachability, and (iv)~collisions in both the idle and the
anticipated arrival pose.  The closest reachable valid stance is preferred.  A
blocked nominal stance can be snapped to a valid point within
$0.60\,\mathrm{m}$; the last 4--6 frames then execute a short
(${\approx}0.10\,\mathrm{m}$) approach to establish the final interaction
distance.

\paragraph{Occupancy grid.}
Routes are planned on a $0.10\,\mathrm{m}$ ground grid rasterized from the same
collision geometry used by manipulation, with separate layers for walls,
furniture, door leaves, and floor support.  Obstacles are included only where
they intersect the human's traversed height band, i.e., $0.20$--$1.90\,\mathrm{m}$
above the local floor for walls and $0.05$--$1.90\,\mathrm{m}$ for furniture;
cells without floor support are blocked.  A Euclidean distance transform
supplies the clearance $D(n)$ of each free cell.  To avoid wrongly declaring
narrow rooms disconnected, the grid is inflated by the largest radius in
$\{0.20,0.15,0.10\}\,\mathrm{m}$ that still preserves a connected, reachable
solution.

\paragraph{Search.}
We run eight-connected A* with diagonal corner cutting disabled and transition
cost
\begin{equation}
 c(n,n') = r\left[
   \ell_8(n,n')
   + \lambda\max\bigl(0,d_{\mathrm{safe}}-D(n')\bigr)
   + c_{\mathrm{door}}(n')
 \right],
\end{equation}
where $r=0.10\,\mathrm{m}$ is the grid resolution,
$\ell_8\in\{1,\sqrt{2}\}$ the step length, $\lambda=8$, and
$d_{\mathrm{safe}}=0.35\,\mathrm{m}$.  The clearance term biases paths toward
open space without turning narrow passages into hard obstacles.  For a closed
door, the planner identifies the blocking leaf, routes to a
collision-validated handle stance, executes the door-opening action, updates
occupancy, and replans; if no such sequence exists, the action returns
\texttt{PATH\_BLOCKED} rather than teleporting the human.  String-pulling
simplifies the polyline only when a shortcut is collision-free and does not
increase the weighted cost.

\paragraph{Execution and 3-D validation.}
The route is converted to an ACCAD gait rendered at 12\,fps with a nominal
speed of $1.1\,\mathrm{m\,s^{-1}}$.  A heading controller uses a
$1.2\,\mathrm{m}$ look-ahead, limits yaw change to $10^{\circ}$ per frame, and
blends toward the final object-facing orientation over the last
$1.5\,\mathrm{m}$.  Translation may pause during large turns, keeping the
offset between body heading and travel direction below ${\approx}35^{\circ}$.
Finally, the articulated body, arm, and egocentric eye volumes are checked
along the generated 3-D trajectory.  If a cell that appeared free in 2-D fails
this check, a temporary $0.20\,\mathrm{m}$ avoidance disk is inserted and the
route is replanned; a bounded repair budget prevents infinite retries and
yields a structured failure when no safe route exists.

\subsubsection{Target-Driven Adaptive Manipulation}
Manipulation is synthesized from a small set of reusable pose prototypes,
target-relative end-effector curves, and online inverse kinematics (IK), rather
than from a stored motion clip per object instance.

\paragraph{Contact and grasp selection.}
The target's oriented bounding box determines the contact region: broad, flat
objects are approached near a reachable perimeter edge, and volumetric objects
at a body- or face-centered target.  Geometry also selects the grasp family:
\textsc{pinch} for small or flat objects, \textsc{cylindrical} when height
exceeds ${\approx}1.15\times$ the cross-sectional diameter, and \textsc{round}
otherwise.  This estimate is considered jointly with stance selection, so
navigation never commits to a stance from which the chosen contact is
unreachable.

\paragraph{Body posture and arm IK.}
The body controller searches an adaptive posture ladder: counter-height targets
keep an upright torso or mild lean; low targets use a
${\sim}40^{\circ}$--$86^{\circ}$ forward lean coupled with hip, knee, and ankle
flexion; and shelf insertion favors leg flexion with a more upright torso.  A
cheap arm-length bound prunes infeasible candidates.  For the remainder,
closed-form cyclic coordinate descent adjusts collar, shoulder, and elbow
rotations for up to 28 iterations, with a nominal $1\,\mathrm{mm}$
end-effector convergence target and action-specific final tolerances.  If no
body/arm solution exists, the action returns \texttt{OUT\_OF\_REACH}; a
geometrically reachable but colliding approach returns
\texttt{TRAJECTORY\_COLLISION}.

\paragraph{Reach and tool trajectories.}
Every reach is expanded into pre-contact, contact, and retreat phases.  The
pre-contact offset is typically $0.10$--$0.18\,\mathrm{m}$, shortened or
redirected according to local geometry, and retreat height increases inside
enclosures.  Shoulder, elbow, wrist, forearm-midpoint, and metacarpal samples
are checked throughout the curve.  Action-specific trajectories are defined in
the target frame: a repeated vertical stroke for cutting, a horizontal circle
for stirring, a raster sweep for wiping, and rotation of the held source along
a target-rim-centered path for pouring.  IK and collision checks run at every
sampled point before frames are committed.

\paragraph{Hand adaptation.}
The selected 45-D hand prototype is adapted to the live mesh by curling each
finger in $6^{\circ}$ increments using batched SMPL+H forward kinematics.  A
finger contact is valid within a $6\,\mathrm{mm}$ surface band with at most
$4\,\mathrm{mm}$ penetration, and a grasp is accepted when at least four of
five fingers make valid contact.  Otherwise, the controller tries wrist
perturbations within a natural range (yaw $\pm10^{\circ}$, $\pm20^{\circ}$;
roll $0^{\circ}$, $\pm15^{\circ}$), scores surface contacts, and resumes finger
bending from the best candidate.  Surface distance is queried from a padded
voxel signed-distance cache in object-local coordinates, which makes the first
encounter with a novel mesh costlier than subsequent grasps but avoids manually
authored collision points.  On successful contact, the simulator stores the
object transform in the wrist frame,
\begin{equation}
  {}^{W}\!T_{O} =
  \left({}^{\mathrm{world}}\!T_{W}\right)^{-1}
  {}^{\mathrm{world}}\!T_{O},
\end{equation}
after which the object follows the wrist rigidly, without an artificial upright
snap, until release.

\paragraph{Pose assets and extensibility.}
Pose data are deliberately compact.  \texttt{hand/atlas/} contains three grip
families, \texttt{side\_wrap\_family.npz}, \texttt{round\_wrap\_family.npz},
and \texttt{pinch\_family.npz}, each storing a 45-D right-hand pose and
optionally a family identifier, grip type, preferred world wrist rotation, and
seat template.  Optional object calibrations in \texttt{hand/atlas/objects/}
currently cover cup, ball, box, plate, pen, and tool templates; they store
build parameters and, when calibrated, an exact object-in-wrist transform.
Objects without an entry derive their seat from the family template and
measured cross-section, after which live finger wrapping adapts the pose to the
instance.  Supporting a new object therefore normally requires only baking its
geometry and metadata, plus, if automatic seating is inadequate, a small
calibration file with the desired wrist-relative transform.  A genuinely new
grip requires one family NPZ (e.g., from the prototype-generation utilities),
registration of its identifier in \texttt{\_PROTO\_FAMILY}, and optionally a
validated \texttt{grasp\_hint} specifying the target region (\texttt{auto},
\texttt{center}, \texttt{edge}, or \texttt{body}), family (\texttt{auto},
\texttt{pinch}, or \texttt{wrap}), edge preference, height fraction, and
clearance; no new full-body animation is needed.  A new manipulation primitive
provides its target-relative curve and declares whether each
affected object is uncoupled, rides the end effector, is joint-driven, or
remains body-anchored.
\subsection{Implementation Details of Semantic Layer}
\label{app:simulator-semantic}

\subsubsection{Unified Action Interface}
The public interface uses a deterministic grammar,
\begin{equation}
  \texttt{[Verb] <entity\_name> (instance\_id)
  [<entity\_name> (instance\_id) ...]}.
\end{equation}
Commands are separated by newlines or comma--space delimiters, with
bracket-aware splitting.  Entity names are normalized to lower-case underscore
form and resolved first by exact key and then case-insensitively; an optional
resolver accepts uniquely grounded descriptions but rejects ambiguous
references.  The registry validates entity arity before compiling each command
into one or more typed goals.  Manipulations automatically insert a
\texttt{NAVIGATE\_TO} goal when needed, select a fresh nearest-reachable stance
at execution time, and deduplicate redundant navigation.  For example, a
pick-and-place request is expressed as
\begin{quote}\ttfamily
  [Walk] <red\_mug> (2)\\
  {}[Grab] <red\_mug> (2)\\
  {}[PutIn] <red\_mug> (2) <drawer> (1)
\end{quote}
The same request may also be given in the structured TaskPack form (fields
\texttt{verb}, \texttt{args}, and optional \texttt{indices},
\texttt{scalar\_arg}, and \texttt{place\_spot}), which is rendered into the same
textual grammar before execution.

\paragraph{Arguments and defaults.}
Entity arguments default to instance index~1, except symbolic connector
arguments of \texttt{attach}, which default to index~0.  The suffix
\texttt{@place\_spot=u:v} selects a normalized target-local location, with
$u\in[-1,1]$ from left to right in the target-facing frame and $v\in[0,1]$ from
front to back.  \texttt{[Forward]} defaults to $0.6\,\mathrm{m}$; a unitless
number denotes paces, an \texttt{m} suffix denotes meters, and displacement is
capped at $20\,\mathrm{m}$.  \texttt{[TurnLeft]} and \texttt{[TurnRight]}
default to $90^{\circ}$ and accept angles in $(0,360]^{\circ}$, whereas
\texttt{[TurnAround]} is fixed at $180^{\circ}$.  \texttt{[Wait]} defaults to
3\,s and accepts durations from 0.1 to 600\,s.

Table~\ref{tab:simulator-actions} lists the 36 canonical primitives in the
authoritative runtime registry, all marked implemented in the audited
revision.  Surface aliases are case-insensitive after normalization and map to
the same primitive.  A task-facing prompt may expose only a relevant subset
without changing the executor's registered capability set.

\begingroup
\small
\setlength{\tabcolsep}{4pt}
\renewcommand{\arraystretch}{1.15}
\begin{longtable}{@{}>{\raggedright\arraybackslash}p{0.20\linewidth}
                   >{\raggedright\arraybackslash}p{0.27\linewidth}
                   >{\raggedright\arraybackslash}p{0.47\linewidth}@{}}
\caption{The 36 canonical action primitives, grouped by function.  The second
column lists the accepted surface verbs (written as \texttt{[Verb]} in a
command, e.g., \texttt{[Walk]}) followed, in italics, by the arguments in
order.  Distance, angle, duration, and angle error are numeric; all other
arguments refer to scene entities.  ``Held'' means that the named object must
match the persistent held-object state.}
\label{tab:simulator-actions}\\
\toprule
Primitive & Surface verbs and arguments & Operational semantics \\
\midrule
\endfirsthead
\caption[]{(continued)}\\
\toprule
Primitive & Surface verbs and arguments & Operational semantics \\
\midrule
\endhead
\midrule
\multicolumn{3}{r}{\footnotesize\textit{Continued on next page}}\\
\endfoot
\bottomrule
\endlastfoot
 
\multicolumn{3}{@{}l}{\textbf{Locomotion and posture}}\\*
\texttt{navigate\_to} &
\texttt{Walk}, \texttt{Run}, \texttt{WalkTowards}\newline
\textit{target} &
Plans and executes a collision-aware route to a live interaction stance near
the target; \texttt{Run} is an alias and selects no separate speed
controller. \\
\texttt{forward} &
\texttt{Forward}, \texttt{WalkForward}\newline
optional \textit{distance} &
Collision-checks and executes a straight body-frame displacement. \\
\texttt{turn} &
\texttt{TurnLeft}, \texttt{TurnRight}, \texttt{TurnAround}\newline
optional \textit{angle} &
Rotates in place while preserving any wrist-carried object; the optional angle
applies to \texttt{TurnLeft} and \texttt{TurnRight} only. \\
\texttt{sit} &
\texttt{Sit}\newline
\textit{seat} &
Lowers and anchors the body on the live seat top; records \textsc{sitting}. \\
\texttt{standup} &
\texttt{StandUp} &
Reverses the seated pose, steps away from the seat, and clears
\textsc{sitting}. \\
\texttt{squat} &
\texttt{Squat}, \texttt{SquatDown} &
Lowers the body and egocentric camera into a crouch, then returns upright. \\
\addlinespace
 
\multicolumn{3}{@{}l}{\textbf{Grasping and placement}}\\*
\texttt{pick} &
\texttt{Grab}\newline
\textit{object} &
Reaches, closes the adaptive hand, and establishes persistent held state;
fails if another object is held. \\
\texttt{release} &
\texttt{Drop}, \texttt{Release} &
Releases the held object at its carried pose, makes it dynamic, and settles
it. \\
\texttt{flip} &
\texttt{Flip}, \texttt{TurnOver}\newline
\textit{object} &
Rotates the named held object by $180^{\circ}$ without releasing it. \\
\texttt{place} &
\texttt{Put}, \texttt{PutBack}, \texttt{PutObjBack}, \texttt{PutOn}\newline
\textit{object, surface} &
Places the named held object on a support, settles physics, and verifies
\textsc{on}. \\
\texttt{place\_inside} &
\texttt{PutIn}, \texttt{PutInside}, \texttt{Insert}\newline
\textit{object, container}, optional \textit{region} &
Places the held object in a named/indexed interior region or an automatically
distributed free region; verifies \textsc{inside}. \\
\texttt{place\_on\_floor} &
\texttt{PutOnFloor}\newline
\textit{object, anchor} &
Crouches, places the named held object on the floor beside the anchor, and
verifies proximity. \\
\texttt{fall\_to\_floor} &
\texttt{FallToFloor}\newline
\textit{object, support anchor} &
Applies a deterministic unattended transition from the stated support to the
adjacent floor; rejects held, contained, or incorrectly supported objects. \\
\addlinespace
 
\multicolumn{3}{@{}l}{\textbf{Assembly}}\\*
\texttt{attach} &
\texttt{Attach}, \texttt{Assemble}, \texttt{Install}\newline
\textit{child, child connector, parent, parent connector}, optional
\textit{angle error} &
Mates declared connectors at their 6-DoF target pose; the optional integer
angle error perturbs the mating orientation for controlled evaluation. \\
\texttt{detach} &
\texttt{Detach}, \texttt{Disassemble}, \texttt{Remove}\newline
\textit{child, parent} &
Removes the child--parent attachment and restores independent manipulation. \\
\texttt{raise} &
\texttt{Raise}, \texttt{Erect}, \texttt{StandUpright}\newline
\textit{assembly} &
Rotates a flat assembly upright about its base, preserving internal
attachments. \\
\addlinespace
 
\multicolumn{3}{@{}l}{\textbf{Fixtures and articulated objects}}\\*
\texttt{push} &
\texttt{Push}, \texttt{PushRack}\newline
\textit{fixture, destination} &
With the hand empty, pushes a registered movable fixture to the destination;
supported contents move with it. \\
\texttt{pull} &
\texttt{Pull}, \texttt{PullRack}\newline
\textit{fixture, destination} &
With the hand empty, pulls a registered movable fixture to the destination,
propagating contained objects. \\
\texttt{open} &
\texttt{Open}\newline
\textit{openable object} &
Drives a registered swing or sliding joint to its open limit; verifies
\textsc{open}. \\
\texttt{close} &
\texttt{Close}\newline
\textit{openable object} &
Drives the articulated joint to its closed limit; verifies \textsc{closed}. \\
\texttt{open\_door} &
\texttt{OpenDoor}\newline
\textit{door} &
Reaches the handle, swings the door, and removes the opened leaf from
navigation occupancy. \\
\texttt{close\_door} &
\texttt{CloseDoor}\newline
\textit{door} &
Closes the hinged door and restores its blocking occupancy in the navigation
grid. \\
\addlinespace
 
\multicolumn{3}{@{}l}{\textbf{Devices and tool use}}\\*
\texttt{turn\_on} &
\texttt{TurnOn}, \texttt{SwitchOn}\newline
\textit{device} &
Physically presses the registered control and establishes \textsc{running}. \\
\texttt{turn\_off} &
\texttt{TurnOff}, \texttt{SwitchOff}\newline
\textit{device} &
Presses the control and clears \textsc{running}. \\
\texttt{cook} &
\texttt{Cook}\newline
\textit{food} &
Requires the food to be in/on a suitable powered appliance, performs the
interaction, and verifies \textsc{cooked}. \\
\texttt{cut} &
\texttt{Cut}\newline
\textit{target, held tool} &
Executes a collision-checked chopping trajectory; verifies \textsc{cut}. \\
\texttt{stir} &
\texttt{Stir}\newline
\textit{target, held tool} &
Executes a circular tool trajectory inside the target; the motion itself is
rendered and logged. \\
\texttt{pour} &
\texttt{Pour}\newline
\textit{target, held source} &
Tilts the held source around the target rim; for configured contents,
verifies transfer into the target. \\
\texttt{wipe} &
\texttt{Wipe}\newline
\textit{tool, target} &
Executes a raster sweep, acquiring the tool if the hand is empty; no separate
cleanliness predicate is scored.  The public order is tool first; the legacy
target--tool order is recognized heuristically. \\
\texttt{use\_computer} &
\texttt{UseComputer}, \texttt{Type}\newline
\textit{keyboard or workstation} &
Optionally auto-sits, types with both hands, activates the display, and
auto-stands if this action introduced the sitting. \\
\texttt{answer\_phone} &
\texttt{AnswerPhone}\newline
\textit{phone} &
Requires the named mobile phone to be held and moves it to the ear without
navigating back to its original location; subsequent waiting preserves this
pose. \\
\addlinespace
 
\multicolumn{3}{@{}l}{\textbf{Perception and control}}\\*
\texttt{look\_around} &
\texttt{LookAround} &
Performs a level, full-circle egocentric scan. \\
\texttt{look\_scan} &
\texttt{LookScan} &
Performs a floor-pitched sweep for nearby or low objects. \\
\texttt{look\_at} &
\texttt{LookAt}\newline
\textit{entity} &
Aims gaze at the entity; may reposition or reorient a label-bearing object to
obtain a readable inspection view. \\
\texttt{wait} &
\texttt{Wait}\newline
optional \textit{duration} &
Holds the current pose and renders stationary frames for the requested
duration. \\
\texttt{confirm} &
\texttt{Confirm} &
Marks a human-policy halt/finalization point; intentionally emits no physical
animation frames. \\
\end{longtable}
\endgroup

\subsubsection{Verifiable Multi-Turn Interaction}
The evaluation involves three participants: an assistant policy, a
human-executor policy, and a persistent Isaac environment.  Their information
flow at turn $t$ is
\begin{equation}
  (I_{\leq t},u^{H}_{t-1},\tau_t)
  \xrightarrow{\text{assistant}} u^{A}_t
  \xrightarrow{\text{human}} (c_t,u^{H}_t)
  \xrightarrow{\text{environment}} (I_{t+1},r_t,e_t),
\end{equation}
where $I$ denotes sampled egocentric frames, $\tau$ the task instruction, $u^A$
and $u^H$ the two agents' utterances, $c_t$ a grounded command, $r_t$ the
observed relation delta, and $e_t$ a structured execution outcome.  The
environment persists across turns: object transforms, articulated joints,
attachments, containment, the human pose, and held-object state are never
reconstructed from language.

\paragraph{Reset and information boundaries.}
At reset, the evaluator loads a frozen task/runtime, resets or reuses a warm
Isaac server, verifies that setup has not already satisfied any scored goal,
executes an authored visual prologue, and pins its evidence-bearing opening
frames.  On each subsequent tick, the assistant receives only the sampled
first-person visual segment, the task instruction, recent human speech, and a
latest-segment marker; internal instance identifiers, relation tables,
coordinates, held state, answer keys, and score are hidden.  It responds with
\texttt{\$silent\$}, an intervention prefixed by \texttt{\$interrupt\$}, an
error report prefixed by \texttt{\$error\$}, or, in evaluation, the strict
completion token \texttt{\$done\$} (equivalently \texttt{<done/>}).

The human executor receives the same visual evidence plus the currently
visible/actionable object handles, held state, the assistant's utterance, the
previous engine outcome, and the command grammar, and returns separate
\texttt{<reasoning>}, \texttt{<action>}, and \texttt{<speak>} fields.  Private
reasoning and commands are logged but never shown to the assistant: only the
human's speech, which is checked for accidental internal-ID leakage, and the
pixels produced by execution cross this boundary.  An action block may queue
several lines, but only one command executes per tick and the assistant is
polled between commands.  In the TaskPack interaction runtime, a new
intervention can preempt pending commands, and a parse or execution error
flushes the remainder so that stale actions are not applied after a changed
instruction.

\paragraph{Command execution and verification.}
Each command passes through five stages: (1)~parsing and grounding under the
fixed registry, recording whether the command was strictly accepted or
repaired; (2)~compilation into typed goals, including implicit navigation;
(3)~execution by the physical controllers described above; (4)~physics
settling and read-back of the actual relations and articulated state; and
(5)~postcondition validation and rendering of the next first-person segment.
A failed primitive returns a structured \texttt{PrimitiveError} and no success
frames.  Even when a controller reports success, state-changing actions (open,
close, power, cut, cook) are re-read from the live physical relations; a
mismatch is rewritten as \texttt{POSTCONDITION\_FAILED} and fed back to the
human in plain language for recovery.

\paragraph{Scoring.}
The goal checker supports the physical predicates \textsc{on},
\textsc{inside}, \textsc{held}, \textsc{open}, \textsc{closed}, \textsc{off},
\textsc{near}, \textsc{sitting}, \textsc{cut}, \textsc{cooked},
\textsc{running}, \textsc{attached}, \textsc{raised}, and \textsc{unmoved}.
Scoring reads the current physical relation set, positive goals, prohibitions,
and successful trace outcomes.  Predicates already true before handoff are
baseline-excluded, so agents receive no credit for the initial state.  An
assistant completion declaration stops the interaction but cannot manufacture
success: the world is scored exactly as it stands.

\paragraph{Authored human policies.}
Authored multi-turn tasks may additionally use a fail-closed, acyclic
\texttt{human-policy/1.0} state machine whose branches depend on timeout, world
state, or the conjunction of assistant semantics and world state.  A frozen
semantic verifier maps assistant speech to an exact JSON tuple of intent,
referents, and facts; the policy advances only if this tuple matches the
authored transition and an independently queried world predicate agrees.
Completion specifications can require alternative or negated physical states,
exact relation counts, successful command multisets or sequences, timely
interventions, and forbidden-before constraints, so persuasive but ungrounded
dialogue cannot satisfy a task.

\paragraph{Logging.}
The evaluator keeps an append-only, crash-safe record of assistant and human
prompts, raw and parsed outputs, silence, grounding/repair provenance, every
attempted primitive and its arguments, actual relation deltas, required facts,
success or error codes, frame intervals, wall time, score, and stop reason.
These records make both physical execution and multi-turn communication
auditable after an episode; a missing measurement is treated as unavailable
rather than silently converted to zero.

The same loop supports the \textsc{Ask}, \textsc{Find}, \textsc{Guide},
\textsc{Watch}, \textsc{Narrate}, and \textsc{Orchestrate} interaction
families.  They differ in which participant initially holds information and
when the assistant should intervene, while sharing the same physical executor,
command registry, trace format, and postcondition-based verifier.

\section{Implementation Details of InteractiveBench}
\label{app:interactivebench}

InteractiveBench evaluates an egocentric assistant in closed loop rather than as a
static video-question-answering model.  In each episode, the evaluated Assistant receives
chronologically ordered first-person-view observations and
natural-language speech; and a separate Human executor converts the
Assistant's utterances into environment actions.  The benchmark comprises 120
frozen episodes in six equally weighted reported families of 20 tasks each.  The unit of evaluation is a versioned
\emph{TaskPack}, which fixes the scene, entities, opening event history, public
utterances, executable action programs, score contract, and provenance hashes.

\subsection{Benchmark Construction}
\label{app:balanced120-construction}

\textbf{Construction.} We construct each InteractiveBench task from a task family and a target capability. We first propose a causal scenario with a concrete goal or error to avoid, a role for the assistant, and actions the Human can perform. Two independent reviews check whether the proposal fits the family, requires assistance, can be enacted in a physical scene, and has a measurable outcome without help. Proposals that fail these checks are revised within a fixed budget or discarded.

Before choosing a scene, we define an assessment contract. It records the question being tested, the evidence the assistant must use, when help is needed, the information gap, the success condition, and the expected outcome without help. This contract remains fixed throughout construction. We then bind the scenario to an available scene, objects, and actions, checking that the chosen assets exist and preserve the intended test. Next, we author the initial world state, the events before the Human's request, the request itself, an oracle action sequence, a no-help sequence, and state-based goal checks. These elements form an executable task specification.

We run the event history and both action sequences through deterministic checks. They test action syntax, object affordances, hand occupancy, geometry, goal states, and whether the task is already complete or its answer is exposed too early. For repairable errors, we revise the scene binding, world state, and task script together, then rerun the checks. If a valid repair would change the assessment contract, we discard the candidate. A final, non-editing review checks that the task still requires the intended evidence, timing, and assistance.

Accepted specifications are compiled into TaskPacks and replayed in Isaac Sim. We execute the oracle, score the resulting physical state, and inspect the rendered views for visual errors. We also test the interaction with scripted and model-based Humans, together with no-help, incorrect-help, and early-help controls. Failures caused by task content lead to a new task revision and another validation pass; simulator or infrastructure failures are fixed and retried without changing the task. Finally, we select 120 validated tasks, with 20 in each of the six families. We freeze the TaskPacks, code, prompts, scoring rules, and per-task revision records for reproducibility.

\begin{figure}
 \centering
 \centerline{\includegraphics[width=\linewidth]{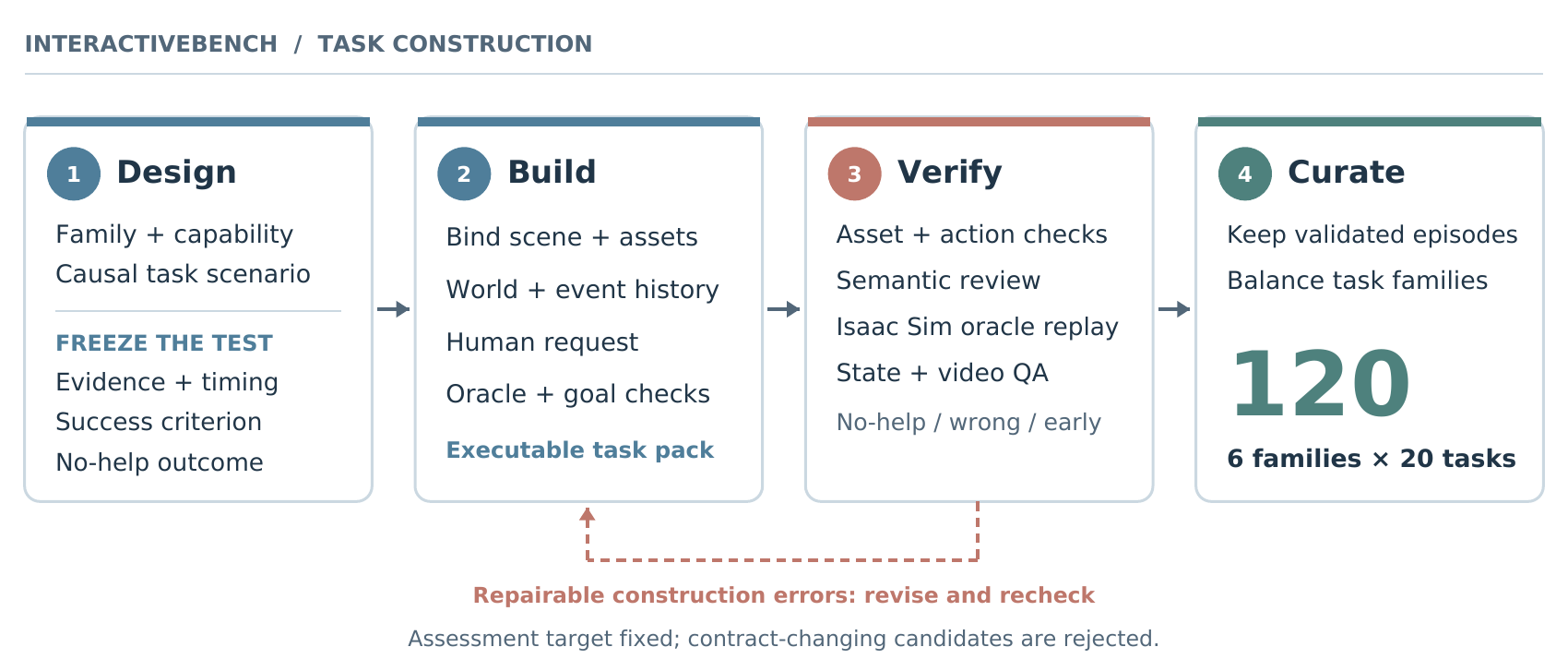}}
 \vspace{-10pt}
\caption{\textbf{InteractiveBench task generation.} For each task family and target capability, we define a scenario and fix its assessment target. We then build an executable task, check its semantics and actions, and validate it through Isaac Sim replay, visual inspection, and causal controls. The dashed arrow shows how repairable errors return for revision and revalidation without changing the assessment target. The final benchmark contains 120 tasks, with 20 per family.}
\vspace{-10pt}
\end{figure}

\subsubsection{Design objective and candidate generation}
 
\paragraph{Design objective.}
Construction targets interactive episodes rather than scripted renders.  In a
valid episode, (i)~task-relevant evidence is genuinely visible from the wearer
camera; (ii)~this evidence can later leave the field of view or be occluded by
intervening activity; (iii)~the wearer asks for help, or proceeds toward a
mistake, at a causally appropriate time; (iv)~the Assistant can change the
outcome through natural-language intervention; and (v)~correct and incorrect
assistance leave different verifiable world states.
Requirement~(v) excludes tasks whose success exists only in prose. 
 
\paragraph{Candidate generation.}
Stories were generated per family.  Up to six family workers ran in parallel,
but generation within a family was serial, so each new proposal saw digests of
previously accepted and rejected same-family proposals.  This
``avoid-from'' context, containing both prior stories and rejection reasons,
discouraged paraphrased duplicates and repeated design failures.  The frozen
authoring campaign used \bcode{gemini 3.1 pro} as author and
\bcode{gpt 5.5} as judge, with random seed~11, deterministic
decoding (temperature~0), two independent review passes, and up to eight author
revisions after rejection.  The reviewer checked family fit, assistant
necessity, physical expressibility, counterfactual discriminability, safety,
and within-family causal diversity, and returned \bcode{KEEP},
\bcode{KEEP_DEBT}, or \bcode{RETRY}.  \bcode{KEEP} admitted a story to
compilation; it was not a benchmark pass.

\subsubsection{Interaction assembly and release freeze}
 
\paragraph{Interaction TaskPack.}
Each physical TaskPack is converted into a derived interaction TaskPack without
changing its parent physical state.  The derived layer adds a natural
\bcode{prelude_say}, deterministic opening actions, a handoff utterance, frame
selection, public and Human-only task contexts, spoken-cue contracts, WATCH
intervention windows, and completion requirements.  Opening actions are
executed in the simulator and shown to the Assistant as pixels, never as a
symbolic textual summary.  The role-based information partition is described
in \S\ref{app:balanced120-info}.
 
\paragraph{Pre-sweep controls.}
Before entering a model sweep, each task's closed-loop assembly was challenged
in a fixed order: (1)~an oracle-verbal Assistant with a scripted Human checked
wiring and action anchors; (2)~the same oracle with the model Human measured the
executor ceiling; (3)~silent, deliberately wrong, and deliberately early
responders checked that the Human cannot solve the task independently and that
failures are observable; and (4)~a strong-model canary exercised free dialogue.
The resulting audit separates eight layers---infrastructure, available
evidence, Assistant decision, Human grounding, process/transition, world state,
completion, and visual quality---so that a physically successful oracle video
is evidence of feasibility, never a model score.

\subsection{Evaluation Protocol}
\label{app:balanced120-evaluation}
 
\subsubsection{Two-agent closed loop}
 
Each evaluation cell pairs one evaluated Assistant with a fixed Human executor
and one simulator instance.  In the documented configuration the Human model is
\bcode{Gemini-3.1-pro}, held constant across compared Assistant models. An
episode proceeds as follows.
 
\begin{enumerate}
  \item \textbf{Load.}  Load the exact TaskPack and verify its digest,
  family-specific interaction specification, scene assets, model route, frame
  policy, and fresh-process configuration.
  \item \textbf{Opening.}  Execute the deterministic opening.  Each successful
  setup command yields a chronological FPV interval; setup failure invalidates
  the cell.  Opening actions are unscored.  The score baseline is captured
  after the opening, so a temporarily satisfied predicate cannot leak free
  credit.
  \item \textbf{Assistant turn.}  At interaction tick $t$, send the Assistant
  the newly selected FPV images, any new wearer speech, and the literal marker
  ``Latest segment.''  The first turn also carries the common system prompt and
  family addendum.  No private symbolic state is included.
  \item \textbf{Parse.}  Parse exactly one Assistant control response.  Speech
  is passed to the Human.  The response \texttt{\$silent\$} yields no speech,
  and \texttt{\$done\$} freezes the world for scoring; the freeze is deferred
  only while an already accepted Human action sequence is draining.
  \item \textbf{Human turn.}  Build a stateless Human turn from the public task
  context, current motor-grounding legend, held object, last action result, new
  Assistant speech, and any still-active instruction.  WATCH additionally
  supplies the Human's private next no-advice action, which is never visible to
  the Assistant.  The Human returns tagged reasoning, zero or more requested
  commands, and optional speech.  The harness executes at most one command per
  simulator tick and polls the Assistant again after the resulting visual
  segment.
  \item \textbf{Log.}  Record the proposed and executed command,
  success/refusal receipt, frames and their provenance, structured visibility,
  held state, semantic state transitions, both agents' raw and parsed outputs,
  and latency.  A rejected action is observable and must be recovered in a
  later turn; it is never silently converted into success.
  \item \textbf{Terminate.}  Continue until the Assistant declares done or the
  authored action/tick budget is exhausted.  Goal satisfaction never
  auto-terminates the episode; otherwise a model could enumerate alternatives
  and be scored at the first lucky state.  The final state and complete trace
  are then evaluated offline.
\end{enumerate}
 
\paragraph{Frame selection.}
The FPV buffer preserves causal temporal structure.  Each simulator command
forms one interval, uniformly sampled to a configured number of frames while
always retaining the true result frame; a \bcode{LookAt} interval additionally
pins a \bcode{look_hold} frame so that the readable cue is not replaced by the
release frame.  The authored opening history is pinned or budgeted as specified
by the TaskPack; later turns normally contribute only the latest interval.
Stateless backends receive a bounded temporal selection that keeps the two most
recent visual turns intact and samples older history uniformly, capped at 32
images in the documented configuration.  All families supply chronological FPV
evidence to the Assistant; NARRATE supplies no images to the Human.
 
\subsubsection{Information boundaries and Human restrictions}
\label{app:balanced120-info}
 
Information is partitioned by role.  The Assistant sees FPV images, wearer
speech, its own dialogue history (or an explicitly declared memory treatment),
and the public task request.  It never sees simulator identifiers, the motor
legend, the answer key, verification points, score, hidden completion state,
oracle actions, or the Human's private WATCH plan.  The Human sees only the
information needed to ground an Assistant instruction to an available action
handle, and is explicitly forbidden to infer the answer, select a destination
from the task semantics, or repair the Assistant's plan.  The following
family-specific restrictions are central to causal attribution.
 
\begin{itemize}
  \item \textbf{ASK:} the Human cannot answer, count, read, calculate, or
  choose; only new Assistant speech may authorize one exact action.  An unseen
  movable object requires an explicitly named search source; otherwise the
  Human asks for a visually grounded target and action.
  \item \textbf{FIND:} the Human cannot infer a remembered location or choose a
  search order.  A movable target handle is exposed only while visible or held,
  so valid retrieval requires Assistant-directed search and visible-before-Grab
  grounding; this prevents semantic-ID auto-navigation from turning memory
  failure into physical success.
  \item \textbf{GUIDE:} the Human has no independent procedure.  It executes
  the first currently actionable semantic step for one object, continues only
  its local motor consequences, and asks for the next step when idle.
  \item \textbf{WATCH:} without a motor instruction, the Human advances exactly
  one command along the private no-advice plan.  A new explicit Assistant
  instruction preempts it; acknowledgements do not.  Authored wearer speech is
  emitted at its fixed action cursor and never authorizes a motor act.
  \item \textbf{NARRATE:} the Human receives no images and cannot invent visual,
  audio, tactile, or other evidence.  Private handles encode only an object or
  destination already identified by the Assistant, and at most one immediate
  action is returned.
  \item \textbf{ORCHESTRATE:} the Human knows the requested outcome but may not
  create, optimize, reorder, or repair the plan.  It executes one strategic
  subtask and its local motor sequence, reports the result, and waits for the
  next subtask.
\end{itemize}
 
\subsubsection{Assistant prompts}
 
The box below reproduces the common Assistant system contract verbatim, apart
from line wrapping.  The user's task arrives through natural opening/handoff
speech; ASK may additionally receive a public task-context paragraph.  Neither
the hidden answer nor a symbolic checklist is inserted.
 
\begin{bprompt}{Evaluated Assistant: common system prompt}
\ttfamily
You are an egocentric AI assistant helping a person through smart glasses.\\
You receive the person's spoken request and first-person video in chronological
order. After every environment action, you receive the new visual segment and
any sentence the person said.\\[2pt]
You cannot see the evaluator's score, hidden completion state, answer key, or
simulator ground truth. Judge progress only from the person's speech and the
first-person video you actually receive.\\[2pt]
Choose exactly one response mode each turn:\\
-- \$silent\$ when speaking now would not help.\\
-- \$interrupt\$<one concise, natural utterance> for an answer, useful next
step, clarification, or timely instruction.\\
-- \$error\$<one concise, natural correction> when the person is currently
making a mistake.\\
-- \$done\$ only when you judge the requested task is complete. The world is
scored immediately when you declare done, and nothing afterwards counts.
\$done\$ must be the entire response: never attach an answer or instruction to
it. Speak first with \$interrupt\$, observe the result, and declare done on a
later turn. Completion requires positive evidence that the person's entire
requested activity has finished. A quiet moment, harmless action, unchanged
frame, or absence of visible error is not completion; use \$silent\$ and keep
watching instead.\\[2pt]
Use ordinary English. Refer to objects by visible description and location,
never simulator ids. Do not expose hidden reasoning. Do not claim that an
unseen or undecidable fact is certain.
\end{bprompt}

\begin{bprompt}{Assistant turn payload (schematic, in wire order)}
\ttfamily
[First turn only: ``Before the setup actions, the person says:
\textless prelude\textgreater'']\\
\textless chronological FPV image blocks selected for this request\textgreater\\
{}[If speech occurred: ``After the setup actions, the person says:
\textless handoff\textgreater'' or ``The person says:
\textless live speech\textgreater'']\\
Latest segment.
\end{bprompt}

\subsubsection{Human executor prompts}
 
The Human prompt is held fixed across Assistant models.  The first box gives
the common operative template; the production prompt adds family-specific
anti-leak and malformed-action clauses with the same semantics.  The second box
shows the dynamically filled Human user turn.
 
\begin{bprompt}{Human executor: common prompt contract}
\ttfamily
You are the person wearing smart glasses and physically acting in the world. A
separate Assistant watches your camera and talks to you. You are cooperative and
execute its instructions faithfully, but you do not know the hidden answer key.
You cannot see the evaluator's score, hidden completion state, or simulator
ground truth; only the Assistant may end the episode.\\[2pt]
Ground every referenced object and destination before acting. Act only when the
Assistant's words, available sensory evidence, and the motor legend identify one
command key. If a reference is absent, ambiguous, visually inconsistent, or cut
off, leave action empty and ask one short clarification. Preserve the requested
spatial relation (inside $\rightarrow$ PutInside; on/onto $\rightarrow$ Put) and
never substitute an object, destination, or motor action.\\[2pt]
A new Assistant instruction supersedes an unfinished old one. Carry one object
at a time; consult the held-object state before Grab. Walk to a named destination
before placement unless the latest receipt says you are already there. Read
success/refusal receipts literally, recover only as instructed, and never claim
completion before every requested clause succeeds. Handle at most one primary
object per response.\\[2pt]
Output exactly, in order:\\
<reasoning>two or three short private sentences</reasoning>\\
<action>zero or more grammar-exact simulator commands, one per line</action>\\
<speak>one short spoken sentence, or empty</speak>
\end{bprompt}
 
\begin{bprompt}{Human turn payload (schematic)}
\ttfamily
TASK CONTEXT: \textless public Human context\textgreater\\
TURN MODE: \textless opening / assistant-new / assistant-active / queued /
watch-default / idle\textgreater\\
{}[WATCH only: private no-advice action plan and next cursor]\\
MOTOR REFERENCES: \textless currently grounded handles and public
descriptions\textgreater\\
YOU ARE HOLDING: \textless object or nothing\textgreater\\
NEW ASSISTANT SPEECH: \textless utterance or none\textgreater\\
ACTIVE INSTRUCTION: \textless unfinished instruction or none\textgreater\\
{}[RESULT OF LAST ACTION: \textless success/refusal and reason\textgreater]\\
PRIVATE ACTION SYNTAX: \textless legal command grammar\textgreater\\
Reply with <reasoning>, then <action>, then <speak>.
\end{bprompt}
 
The action language comprises navigation (Walk, Run, Forward, Turn, Look),
manipulation (Grab, Drop, Put, PutInside, Push, Pull, Attach, Detach),
open/close and power operations, and task actions such as Cut, Stir, Pour,
Wipe, Cook, and AnswerPhone.  Named arguments must use the exact current handle
syntax.  The harness extracts commands only from the \bcode{<action>} block,
logs all proposed commands, and executes them one at a time.
 
\subsubsection{Validity and causal attribution}
 
Validity is audited before task scoring.  A valid cell requires a successful
scene/setup, the intended TaskPack hash and model route, a complete interaction
trace, usable frames, and functioning Assistant, Human, transport, and
simulator components.  Infrastructure termination, corrupted or missing
evidence, failed setup, or material Human answer leakage makes a cell
\emph{VOID}, to be rerun.  Assistant silence, wrong guidance, loops, premature
\texttt{\$done\$}, failure to declare, or incomplete work in an otherwise sound
run is a valid model failure.  Minor executor formatting repairs are logged;
Human- or simulator-caused outcomes are never credited to the Assistant.
 
For each selected cell, the audit bundle retains the TaskPack and hash,
\bcode{dialogue.json}, \bcode{eval.json}, the final manifest (or an equivalent
terminal snapshot), all Assistant turns, \bcode{human_action_trace}, action
receipts, structured visibility, opening and post-prologue baselines, final
relations and persistent states, the held object, the stop reason, and the
selected-attempt manifest.  Attempt selection follows a fixed
validity/provenance policy, never score.
 
\subsection{Evaluation Metrics}
\label{app:balanced120-metrics}
 
\subsubsection{Task-level metrics}
 
For a valid task $t$, the source-family evaluator returns strict completion
$s_t\in\{0,1\}$ and a task-normalized completion value $c_t\in[0,1]$.  For a
reported family $F$ of 20 tasks,
\begin{equation}
  \operatorname{PassRate}(F)=\frac{1}{20}\sum_{t\in F}s_t,
  \qquad
  \operatorname{MCR}(F)=\frac{1}{20}\sum_{t\in F}c_t.
  \label{eq:family-metrics}
\end{equation}
MCR denotes Mean Completion Rate; it is not an action ratio.  Each task
is normalized before receiving equal weight, so tasks with many objects,
verification points, bundles, or capability atoms do not dominate.

\subsubsection{Capability-level metrics}
\label{app:balanced120-capability}

Task outcomes and capability diagnoses are scored independently, using the
frozen C1--C9 ontology in Table~\ref{tab:capability-ontology}.
Each task is annotated with its required capabilities, and each capability
is assessed through separate evaluation points, referred to as capability
atoms. A task may contain multiple evaluation points for the same capability.

\begin{table}[ht]
\centering
\small
\setlength{\tabcolsep}{3.2pt}
\caption{Capability ontology.}
\label{tab:capability-ontology}
\begin{tabular}{@{}c>{\raggedright\arraybackslash}p{0.25\textwidth}>{\raggedright\arraybackslash}p{0.63\textwidth}@{}}
\toprule
\textbf{ID} & \textbf{Capability} & \textbf{Boundary} \\
\midrule
C1 & Object/state recognition & Identify the current object and its current state; temporal updating belongs to C4. \\
C2 & Spatial grounding & Locate objects, destinations, and relations in the egocentric scene. \\
C3 & Scene-text reading & Transcribe and bind visible text; conclusions inferred from text belong to C6. \\
C4 & Episodic memory/state tracking & Retain and update identities, events, locations, and states across time. \\
C5 & Planning/replanning & Form, sequence, and revise a plan; Human execution cannot substitute for Assistant planning. \\
C6 & Knowledge/reasoning & Apply rules, arithmetic, domain knowledge, and constraints to choose or infer. \\
C7 & Error/risk detection & Recognize an error, omission, unsafe state, or impending risk. \\
C8 & Intervention timing & Place the first task-relevant intervention between the frozen earliest-valid point and deadline; semantic correctness is scored elsewhere. \\
C9 & Actionable communication & Produce unambiguous, parameter-complete, correctly ordered, physically executable language. \\
\bottomrule
\end{tabular}
\end{table}

\paragraph{Evaluation-point scoring.}
Each required evaluation point receives exactly one label:
\bcode{clean_pass}, \bcode{wrong}, \bcode{abstain},
\bcode{unobserved}, \bcode{assisted}, or \bcode{void}.
Only \bcode{clean_pass} counts as a passed evaluation point.
Required points missed because of Assistant silence, premature stopping,
or missed opportunities are labeled \bcode{unobserved}, not VOID.
These points remain in the denominator and contribute no passes.

\paragraph{Equal-weight aggregation over evaluation points.}
Scores are computed separately for each evaluated model.
For capability $k$, let $E_k$ denote the set of task evaluations with
complete required-point evidence. For each task $t \in E_k$, let
$n_{t,k}$ be the number of required evaluation points for capability $k$,
and let $p_{t,k}$ be the number labeled \bcode{clean_pass}.
The capability score is
\begin{equation}
\operatorname{CapScore}(k)
=
100 \times
\frac{\sum_{t \in E_k} p_{t,k}}
     {\sum_{t \in E_k} n_{t,k}}.
\label{eq:capability-point-pooling}
\end{equation}
Thus, we pool evaluation points across all included tasks and give
each point equal weight. Tasks with more required points contribute
proportionally more to the score; we do not average per-task pass rates.
All capability scores are reported as percentages.

\paragraph{Independence from task completion.}
Capability points are scored using their own checks rather than the
overall task outcome. Whole-task success is not propagated to all
capabilities, nor is whole-task failure propagated as blanket capability
failure. A failed task can therefore contain passed capability points,
while task completion alone does not establish that every required
capability was demonstrated.

\subsubsection{Efficiency and reproducibility}
 
\paragraph{Efficiency.}
Efficiency is computed only on strict-success episodes, against a task-matched
shortest legal oracle.  Each Human macro action (Walk, Look, Open, Grab, Put,
etc.) counts as one step, including refused actions, which consume an
interaction; speech, frames, the deterministic opening, and post-completion
cleanup do not count.  The stable-completion index is the earliest action after
which all strict core conditions hold and remain true.  Ratios are paired per
task before aggregation; dividing a model's mean on one success set by an
oracle mean on another is invalid.  Efficiency is reported alongside, never
folded into, MCR or Pass Rate.
 
\paragraph{Reproducibility.}
For each model--task cell, the released manifest preserves the fields
\bcode{source_family} and \bcode{assigned_family}, the expected and observed
TaskPack hashes, the hash origin, the rubric/version, the completion protocol,
the evaluation status, and the evidence path.  These fields ensure that a reclassification cannot silently
change the scorer and that a run-level observed hash is not conflated with a
frozen-rubric fallback hash.  All aggregate tables are recomputable from the
per-task and per-atom ledgers.
 
\subsection{Task Catalog}
\label{app:balanced120-tasks}

Tables~\ref{tab:guidance-tasks}--\ref{tab:finding-tasks} list every released
task, stating (i)~what the wearer and Assistant must accomplish, (ii)~the
capability the episode isolates, and (iii)~what the source-family evaluator
verifies.  ``Baseline'' denotes the state at the Assistant handoff, after the
deterministic opening, not necessarily the state at simulator reset.

\subsubsection{Guidance (20 tasks)}

Guidance tests whether the Assistant can turn visible documents, domain
knowledge, earlier observations, and numerical constraints into grounded,
incremental instruction. 

\begin{scriptsize}
\setlength{\tabcolsep}{3.2pt}
\begin{longtable}{@{}>{\centering\arraybackslash}p{0.05\textwidth}>{\raggedright\arraybackslash}p{0.42\textwidth}>{\raggedright\arraybackslash}p{0.47\textwidth}@{}}
\caption{Guidance tasks.}\label{tab:guidance-tasks}\\
\bTaskHead{Episode and capability tested}
1 & Read seven beverage labels and solve a globally coupled three-box allocation; tests OCR, arithmetic, and global constraint satisfaction. & Exact composites: A = milk-500 + water-100; B = milk-500 + juice-200; C = milk-500 + juice-200 + water-100.  No wrong categories; all seven bottles form one exact partition. \\
2 & Recall carton sizes seen before a phone interruption and fill three later-arriving orders; tests episodic visual memory and numerical allocation. & Exact size composites A = 500+100, B = 500+250, C = 500+250+100; all seven cartons partitioned once, no wrong sizes, and the phone preserved. \\
3 & Archive obsolete briefing material before placing a new packet and pen; tests document-grounded ordered guidance. & Old material archived; new packet and pen on the reset table; new-item credit depends on removal of the obsolete material. \\
4 & Use a stocking rule and price cards to arrange bread, candy, chocolate, and their labels in ascending-price order. & Three exact two-object composites occupy the low/mid/high promotional slots; none of the six objects remains on the cart. \\
5 & Follow a tomato recipe through cutting, cooking, plating, and serving. & Tomato cut and safely cooked; cooktop off; the cooked form is the sole plated food; completed plate held; knife and spatula returned. \\
6 & Sort four pantry goods by storage requirements; tests rule reading and storage knowledge. & Candy/chocolate exactly in the light display; coriander/coffee exactly in the closed dark cabinet. \\
7 & Choose safe cooling methods for hot lemonade, pur\'ee, and an overheated phone. & Both foods refrigerated with fridge closed; phone on cooling counter with fan running; phone excluded from fridge. \\
8 & Dry a wet towel and boot with material-appropriate equipment. & Towel alone in a closed, running dryer cycle; boot on or in the shoe rack and not in the dryer. \\
9 & Apply a printed, tool-specific cleaning rule to six objects. & Mug and two bottles cleaned with blue sponge; camera/phone with yellow cloth; these five returned to the clean table; paper remains dry/unwiped and is filed. \\
10 & Repair a three-object cyclic misplacement from a reset map. & Exact partition: phone in secure locker, water in fridge, folder on desk; ordinary door state is not independently scored. \\
11 & Stock a capacity-limited shelf from three equal 800-mL cartons without exceeding capacity. & Exactly two cartons moved to the drink shelf and one remains at source; pre-existing milk/orange stock preserved. \\
12 & Resume an interrupted breakfast plan after the ham stage. & Scone inserted and cooked, oven then closed; previously completed ham placement and burner-off state preserved. \\
13 & Pack an IT return kit in dependency order. & Pad returned before headset; phone inside closed secure locker; exact box contents are pad/headset and exact locker content is phone. \\
14 & Prepare a visitor table while respecting privacy/security ordering. & Staff cabinet closed; sign-in sheet, pen, badges on visitor table; spare folder in closed return locker; setup depends on privacy closure. \\
15 & Resume an interrupted four-step afternoon-service plan after two completed steps. & Apple on setup table and clean mug on the vacated first supply table; completed water/towel states preserved; targets exact. \\
16 & Resume a five-step hotel reset after the first two items. & Towel, soap, and toilet roll on bathroom shelf; already completed folder/water destinations preserved. \\
17 & Close a tea station under storage and hygiene rules. & Milk refrigerated, tea dry-stored, paper discarded; station cleaned and cloth returned; mug placed only on the cleaned station with exact service contents. \\
18 & Clear a cooking station and make it operational. & Used mug, paper bag, and bottle cleared to prescribed destinations; pan on cooktop and cooktop running, with activation gated by cleanup and pan placement. \\
19 & Clear clutter, prepare and cook two vegetables, plate, and deliver a stir-fry. & Cleanup complete; both vegetables cut and cooked in the pan; cooktop off; exact finished contents plated at service table; preparation/cooking dependencies hold. \\
20 & Select an exact one-litre, three-category tasting order from six labeled bottles. & Kit contains exactly GIN-200 mL, RUM-300 mL, and WINE-500 mL; the three larger decoys are excluded. \\
\end{longtable}
\end{scriptsize}

\subsubsection{Planning (20 tasks)}

Planning tests strategic decomposition, resource and route optimization,
constraint satisfaction, and replanning after interruption or state change.

\begin{scriptsize}
\setlength{\tabcolsep}{3.2pt}
\begin{longtable}{@{}>{\centering\arraybackslash}p{0.05\textwidth}>{\raggedright\arraybackslash}p{0.42\textwidth}>{\raggedright\arraybackslash}p{0.47\textwidth}@{}}
\caption{Planning tasks.}\label{tab:planning-tasks}\\
\bTaskHead{Episode and capability tested}
1 & Pack an exact shipment from nine capacity-labeled packages; tests numerical planning and subset selection. & Exact water set \{3 L, 3 L, 2 L\}, juice set \{4 L, 3 L, 2 L\}, and 3-L lemonade; seven candidates exactly, excluding water-4 L and juice-5 L. \\
2 & Execute a complete lunch order with schnitzel, broccolini, sealed sauce, tools, cooking, plating, and delivery. & Both foods cooked/browned and plated; cooktop off and spatula returned; exact two-food plate plus sealed sauce delivered to service pass. \\
3 & Execute an ordered, multi-room housekeeping closeout. & Gift inside closed locker; water refrigerated; towel inside washer; washer closed and running with the towel. \\
4 & Prepare a snack and replan when cutting reveals mold. & Moldy bread cut then discarded; fresh bread cut and plated with cheese/apple; safe plate at service pass; negative/exact constraints and knife return hold. \\
5 & Balance three classroom desks with one water and one juice each using a short one-hand route. & Exact water counts across all three desks form one bundle and exact juice counts another; route length is reported separately. \\
6 & Repair a three-way grocery misplacement with a continuous short route. & Rice to grains, barbecue sauce to sauces, and iced tea to beverages, one complete active-result bundle each. \\
7 & Schedule rice and water before heat, then reset ingredients and tools. & Valid cooked porridge in pot on burner; rice bag in closed cabinet; knife in block; burner off; persistent recipe state records correct ingredient order. \\
8 & Use a house map to plan folder-to-bed, phone-to-dining-table, and mug-to-kitchen moves with one necessary empty leg. & Three destination bundles complete; shortest-route quality is a separate efficiency measure. \\
9 & Select, cut, plate, and deliver two allergy/low-sugar constrained fruit plates. & Six cut-fruit bundles on correct plates and two complete plate-delivery bundles; cheesecake, grape, and pineapple remain intact at baseline. \\
10 & Minimize makespan for potato, pear, trout, and bread using one oven. & Four foods cooked on their own plates at the service table; pear cut; oven off and closed. \\
11 & Parallel-schedule pork knuckle, trout, potato, and pear over oven and stove. & Four complete cooked own-plate dishes returned to prep table; pear cut; oven off/closed and stove off. \\
12 & Detect an incorrectly stored cold brew, repair it, and complete a kitchen reset. & Cold brew and pineapple juice in fridge, mug in dishwasher, sponge in sink-side drying zone; fridge and cleaning cabinet closed. \\
13 & Select the maximum feasible four of nine dishes under 30 minutes, one oven, and one stove. & Complete cooked plated baguette, pear, zucchini, and steak; pear cut; five unselected dishes unchanged; appliances safely reset. \\
14 & Plan a short reset route for four objects across a five-room map. & Mug to kitchen, towel to bathroom, parcel to entry, and folder to living room, each as one endpoint bundle. \\
15 & Filter nine dishes by four allergies, then maximize safe output under time/equipment constraints. & Complete cooked plated chicken, pear, zucchini, and steak; pear cut; all unsafe/unselected dishes unchanged; appliances safely reset. \\
16 & Correct a scripted hotel-checkout mistake and finish a four-object reset. & Folder to desk, clipboard to wardrobe shelf, welcome box to entry console, and rinse mug to bathroom shelf; opening misplacement must actually occur. \\
17 & Combine two delivery chains across a twelve-zone office map. & Folder to meeting room, agenda to copy room, package to reception, towel to bathroom, and phone to reception; one bundle per endpoint. \\
18 & Infer seasoning order and coordinate chicken/bread roasting. & Persistent chicken state exactly soy-then-vinegar; complete chicken and bread plates at service counter; oven off and closed. \\
19 & Recover a spoken plan after work and a phone interruption. & Phone returned from dining sideboard to living-room table; towel moved to kitchen; incident folder moved to the other living room; opening interference occurred. \\
20 & Replan when a late spill appears during close-down. & Initial grime removed with sponge; apple on return cart; post-apple spill removed with towel and table dry; both tools returned. \\
\end{longtable}
\end{scriptsize}

\subsubsection{Accessibility (20 tasks)}

Accessibility models a blind or low-vision wearer.  The Human executor receives
no camera images, so the Assistant must provide timely, body-anchored,
goal-relevant visual information.  All tasks use CFSS-StateScore; ``STG'' in
the table refers to the critical transition gates reported as STG-CFSS.

\begin{scriptsize}
\setlength{\tabcolsep}{3.2pt}
\begin{longtable}{@{}>{\centering\arraybackslash}p{0.05\textwidth}>{\raggedright\arraybackslash}p{0.42\textwidth}>{\raggedright\arraybackslash}p{0.47\textwidth}@{}}
\caption{Accessibility tasks.}\label{tab:accessibility-tasks}\\
\bTaskHead{Episode and capability tested}
1 & Visually identify orange juice among a can, carton, and bottle and give body-anchored guidance. & Juice on breakfast table; soda and water remain at baseline (and never move under the transition gate). \\
2 & Repeatedly reassess a growing gap between a pot and bowl, moving a scale only when it fits. & Scale on clearing counter; three bottles on holding counter; pot/bowl at baseline.  STG keeps the scale at source until all bottles are cleared. \\
3 & Compare remembered and current table settings and restore only a moved blue bowl. & Bowl restored to guest table; added coffee cup, unchanged wineglass/water, and all setting objects remain at handoff baseline. \\
4 & Interrupt selection of a hazardous cleanser and redirect the wearer to baby oil. & Baby oil on side table; cleanser stably returned to cabinet, not held/on floor; bath salt at baseline.  STG requires cleanser return before oil acquisition. \\
5 & Admit that spray is absent from the current view, avoid a cactus, scan, and relocate the spray. & Spray on front cabinet; cactus, skin cream, and sponge at baseline; protected objects never move. \\
6 & Prioritize putting down a carving knife safely before searching for a ringing phone. & Knife stably on prep counter and phone held.  STG requires knife safety before leaving or acquiring the phone. \\
7 & Read a \textsc{No Glass} sign, prevent wrong disposal, and locate the glass-return counter. & Bottle stably on return table, neither held nor in waste/on floor.  STG forbids an unsafe intermediate disposal. \\
8 & Filter nearby items by the semantic class ``food,'' then compare distance to leaking deicer. & Apple in living-room waste bin; deicer, sponge, and salad at baseline. \\
9 & Identify two food items among a non-food distractor and guide two refrigerator trips. & Apple and sandwich in closed fridge; sponge at baseline. \\
10 & Detect an out-of-view mug/folder swap and distinguish it from wearer-caused changes. & Mug and folder restored to their respective baseline supports; phone and completed errand sheet preserved. \\
11 & Track water-bottle counts across three visually similar tables. & Exact counts A=3, B=3, C=3, independent of bottle identity; STG prevents any table from exceeding three. \\
12 & Detect broken-glass danger before placing a new blue cup. & New cup on hazard table; broken glass contained and broom returned; original cups unchanged.  STG clears hazard before cup placement. \\
13 & Select safe, genuinely open storage locations in a cluttered kitchen. & Apple in fridge or open counter; water in either; blue mug on open counter; fridge closed; six clutter objects preserved. \\
14 & Read a cleanup guide and execute nested hazardous-waste disposal and ordinary cleanup. & Guide returned; glass inside yellow bag and bag inside red bin; apple/milk refrigerated; cup/bowl at sink; sponge unchanged.  STG enforces bag-before-bin and hazard-first. \\
15 & Remember a handwritten, three-item, cross-room note while ignoring distractors. & Apple in closed fridge, teapot on tea table, blue mug on side counter; white cup and sponge at baseline. \\
16 & Search three laboratory regions for a phone, folder, and beaker. & Phone on instructor desk; folder and beaker each in either allowed native cabinet; coffee, sponge, marker, and water preserved. \\
17 & Correct three sequential off-ticket picks while completing a catering ticket. & Apple, water, and sandwich on conference table; all eight off-ticket items restored.  Intended STG returns each wrong pick before progress. \\
18 & Inspect a refrigerator, detect a shoe hazard, and store milk safely. & Shoe on an allowed stable support outside fridge; milk inside; fridge closed.  STG stages milk, removes shoe first, and closes last. \\
19 & Recall where a phone was left after a long intervening kitchen activity. & Phone on the designated living-room coffee table; source-memory grounding is evaluated separately because endpoint state alone cannot prove recall. \\
20 & Infer the remaining checklist items after two have already been completed. & Phone on office desk and badge on lobby desk; completed folder/mug endpoints preserved. \\
\end{longtable}
\end{scriptsize}

\subsubsection{Decision (20 tasks)}

Decision poses a concrete situated question whose answer has a physical
consequence.  It tests OCR, multimodal evidence integration, arithmetic,
commonsense, state reconstruction, and the ability to turn an answer into an
explicit action rather than relying on the Human to infer it.

\begin{scriptsize}
\setlength{\tabcolsep}{3.2pt}
\begin{longtable}{@{}>{\centering\arraybackslash}p{0.05\textwidth}>{\raggedright\arraybackslash}p{0.42\textwidth}>{\raggedright\arraybackslash}p{0.47\textwidth}@{}}
\caption{Decision tasks.}\label{tab:decision-tasks}\\
\bTaskHead{Episode and capability tested}
1 & Choose the highest-calorie option without exceeding the corrected daily allowance; tests report reading, spoken correction, and constrained argmax ($2000-1780-110=110$). & The 45-kcal chocolate syrup is on the breakfast table. \\
2 & Fuse a children table and gifts-per-child table across rooms; compute $1\times2+2\times1.5+0\times1=5$. & Five gifts distributed exactly 2/2/1 among three cubbies; spare gift remains at supply. \\
3 & Intersect an earlier remembered promo-brand set with twelve later router cartons. & AXO, BYT, and CUE cartons on display; all nine regular cartons at source/unmoved. \\
4 & Reconstruct which items remain unfinished in a seven-point closeout after a long action chain. & Cabinet closed, oven off, mug/plate in cabinet, juice in fridge, can in trash, fridge closed. \\
5 & Count stocked cartons, then consult a price sheet: $5\times\$3+4\times\$4+2\times\$5=\$41$. & \$41 invoice on approval desk; \$39 and \$45 alternatives remain at source/unmoved. \\
6 & Aggregate two cabinet inventories, subtract removed items, and compute deficits against 4 towels, 5 mugs, and 7 waters. & Two replacement towels and one mug inserted; removed towel/water remain staged; both cabinets closed. \\
7 & Bind an earlier Unit-B projector identity to a later A/B/C deployment table. & Unit-B projector on the specified pedestal; Unit-A/C distractors remain at source. \\
8 & Combine two stock tables and replan after observing an occupied primary shelf. & Cat food reaches the DOG-OLD checkout counter; occupying dog-food can remains on its original shelf. \\
9 & Recall which worktable item required immediate refrigeration and apply food-safety knowledge. & Raw trout in closed fridge; pasta and iced tea remain on worktable. \\
10 & Identify the ingredient that makes dough rise from recipe context and functional knowledge. & Yeast on dough table; jam, bowl, flour, and water remain at sources. \\
11 & Select a fruit for enzymatic tenderizing; distinguishes pineapple/bromelain from lemon acidity. & Pineapple on steak counter; lemon, steak, and bowl remain at source. \\
12 & Recall a quantity of 40 and compare case sizes/prices: $5\times104=520<8\times70=560$. & Quote B on approval desk; Quote A remains at source. \\
13 & Recognize the appropriate exercise apparatus and a safe spatial relation for a backpack. & Backpack released on floor beside parallel bars, not on the apparatus. \\
14 & Choose the largest feasible snack under a remembered total-weight limit. & 250-g cookie at lunch staging; coffee remains there; infeasible 351-g cracker stays at source. \\
15 & Generalize an earlier category-level disposition rule to concrete stock items. & Milk and cheese in public trash; apple and cereal remain at source. \\
16 & Add times across rooms and compare completion: option A ends 8:00, B ends 7:55. & Card B on bedroom cabinet; card A remains at source. \\
17 & Subtract bookings from capacities across two sheets and choose the larger remainder (A=16, B=18). & Sheet B on bar; sheet A remains at source. \\
18 & Use visual affordance and functional knowledge to choose a passive room fragrance. & Reed diffuser on entryway cabinet; atomizer and sponge remain at source. \\
19 & Maintain an event ledger without looking back ($+1+1-1+1=2$ mugs), then complete a three-mug setup. & All three mugs on beverage station; sponge and phone at their required baseline locations. \\
20 & Monitor a five-item grocery checklist across a phone interruption and detect that yogurt and feta were omitted. & All five listed items inside closed fridge; full-state check prevents repair from displacing already completed items 1/3/5. \\
\end{longtable}
\end{scriptsize}

\subsubsection{Prevention (20 tasks)}

In Prevention, the wearer follows a private no-advice plan that is mostly
correct but contains one or more consequential errors.  The Assistant should
remain quiet during correct work and intervene before each error's deadline.
All tasks use WATCH-binary; intervention correctness and timing are measured
separately by C7/C8.

\begin{scriptsize}
\setlength{\tabcolsep}{3.2pt}
\begin{longtable}{@{}>{\centering\arraybackslash}p{0.05\textwidth}>{\raggedright\arraybackslash}p{0.42\textwidth}>{\raggedright\arraybackslash}p{0.47\textwidth}@{}}
\caption{Prevention tasks.}\label{tab:prevention-tasks}\\
\bTaskHead{Error and capability tested}
1 & Cold cuts are moved before sanitizer is stored and the chemical cabinet closed; tests procedural order and food/chemical safety. & Mug in dish rack; towel in used sink; sanitizer in closed chemical cabinet; cold cuts on prep table and clean. \\
2 & Adding another two-serving box would make seven, not exactly six; tests quantity tracking after a delayed update. & One- and three-serving boxes plus exactly one interchangeable two-serving box on service table; the other two-serving box at supply. \\
3 & A used sponge remains on the table after the wearer reaches the break area; tests delayed omitted-step detection. & Sponge, mug, and towel in sink; phone on shelf; table physically clean. \\
4 & A third wine and, later, chips violate ``two wines, burger, no sides''; requires two interventions. & Exactly two interchangeable wines and burger on packing table; remaining wine and chips at stock. \\
5 & Dough is about to enter the oven without an X score; tests prerequisite detection and recovery. & Dough cut and in oven; oven closed/running; current bake cycle began only after valid preparation. \\
6 & A used sponge is confused with an identical clean sponge; tests event-bound identity. & Used sponge in used bin; clean sponge in closed clean cabinet; persistent used/clean attributes agree with history. \\
7 & The fridge remains open while the wearer leaves to retrieve a phone; tests off-screen persistent-state monitoring. & Fridge closed; milk, raspberries, and phone all on breakfast table. \\
8 & An oven door remains open when the wearer is about to start it; tests off-screen appliance state. & Pie in oven; door closed; oven running under a safe current cycle. \\
9 & A towel previously used on a table is selected to dry a clean cup; tests instance provenance and hygiene. & Used towel at used station; clean towel returned to counter; cup clean and dry. \\
10 & Before departure, a laptop and open window/fridge remain unresolved although the oven was already checked. & Laptop on departure console; window, fridge, and oven closed. \\
11 & Among five similar towels, one used earlier is selected for a clean plate; tests long-horizon instance partition. & All five towels in their designated clean/used partitions; plate clean and dry. \\
12 & A kitchen faucet remains on while the wearer leaves to answer a phone; tests persistent hazard tracking. & Sink off; phone on call table; rinsed apple on apple table; orange in fridge; plate in dishwasher; fridge closed. \\
13 & Dough is loaded before the required drip tray; tests procedural prerequisite and recovery. & Tray in oven; dough properly supported by tray; oven closed; dough cooked under a valid cycle. \\
14 & Mug and water are placed but a gift is omitted before checklist confirmation. & Mug, water, and gift all on the specified shelf. \\
15 & An extra third scone and later an unlisted donut are added; requires repeated exact-order monitoring. & Exactly two interchangeable scones plus croissant, muffin, and brownie in box; remaining scone and donut at stock. \\
16 & A folded towel is omitted from a guest-room checklist. & Water/folder on desk, towel in closed cabinet, mug on bedside table. \\
17 & The wearer begins wiping before moving a phone and room key; tests order-sensitive cleaning. & Mug/towel at destinations; phone/key on personal table; sponge on cleaning shelf; desk cleaned after obstructions were removed. \\
18 & A gift is placed before the desk is cleaned; tests order-sensitive housekeeping. & Mug, towel, and welcome card at destinations; desk clean; sponge on shelf; gift finally on desk. \\
19 & An apple is taken to a prep table rather than the refrigerator; tests wrong-destination correction and checklist completion. & Folder at records counter; sanitizer in closed cleaning cabinet; apple in closed fridge; checklist at sign-off table. \\
20 & A washer is about to start while a loose sock remains on a shelf; tests omitted-object and cycle integrity. & Sock and towel in closed/running washer; current cycle's frozen contents include both. \\
\end{longtable}
\end{scriptsize}

\subsubsection{Finding (20 tasks)}

Finding tests episodic object memory and evidence-grounded retrieval.

\begin{scriptsize}
\setlength{\tabcolsep}{3.2pt}
\begin{longtable}{@{}>{\centering\arraybackslash}p{0.05\textwidth}>{\raggedright\arraybackslash}p{0.42\textwidth}>{\raggedright\arraybackslash}p{0.47\textwidth}@{}}
\caption{Finding tasks.}\label{tab:finding-tasks}\\
\bTaskHead{Episode and capability tested}
1 & Inspect labels to distinguish freezer from chiller, then classify five cold-delivery items. & Ice cream/lolly exactly in closed freezer; milk, yogurt, and lettuce exactly in closed chiller (GUIDE StateScore-v2). \\
2 & Inspect contents of two otherwise ambiguous refrigerators to infer the dairy unit. & Cheese in secured dairy fridge; existing milk remains there, chicken in the other unit, and the other fridge remains closed (GUIDE StateScore-v2). \\
3 & Track which of two identical glasses received vodka across a phone delay. & Alcohol-free glass on guest table; actual vodka-containing glass on adult counter, using persistent contents rather than hidden identity (WATCH binary). \\
4 & Search three rooms for a sponge hidden in one of several similar low cabinets when its initial location was never shown. & Search exposes sponge before grab; sponge reaches result counter without hidden-ID navigation. \\
5 & Retrieve the one of two identical erasers that touched a loose sheet rather than a folder. & Paper-contact eraser, not distractor, visibly acquired and placed in storage cabinet. \\
6 & Recall the independent locations of two identical coconut-milk bottles. & Garden and fridge bottles directly retrieved to final tray; fridge opened before the hidden bottle is acquired. \\
7 & Recall where sesame oil remained after another sauce was moved and a long distraction followed. & Sesame oil, not the other bottle, placed in target cabinet after visible acquisition. \\
8 & Distinguish the gift left in the living room from an identical gift already visible in the kitchen. & Missing living-room gift, not visible distractor, retrieved to the kitchen result surface. \\
9 & Recall mayonnaise on a counter versus similar jam and return it to the refrigerator. & Mayonnaise visible before grab and placed in opened fridge; jam not substituted. \\
10 & Recall which of two similar bedside cabinets contains a thermometer. & Correct (left) cabinet opened; thermometer becomes visible and reaches infirmary desk; drink/right cabinet not substituted. \\
11 & Search a bank of identical low cabinets for a lunch box whose initial cabinet is unknown. & Correct cabinet genuinely opened and lunch visible before grab; lunch placed on result worktop. \\
12 & Track a mushroom inside a lidded portable box that was moved into a refrigerator. & Fridge opened, box retrieved, lid removed, mushroom independently extracted; mushroom and box both placed on dining table. \\
13 & Recover the first and second of three identical milk cartons after different trajectories. & First and second cartons, not finished third carton, directly retrieved to the fixed work table. \\
14 & Detect that the wearer searched a similar cabinet in the wrong room for a wallet. & Kitchen cabinet opened; wallet visibly retrieved to result counter; phone/wrong cabinet not substituted. \\
15 & Retrieve the text-defined \textsc{Equipment Storage} form rather than the similar \textsc{Food Storage} form. & Equipment form visibly retrieved from opened file cabinet and placed with the food form on break-room table. \\
16 & Predict and rediscover an apple that fell off-screen from a crowded table. & Floor apple visually rediscovered before grab and placed on result counter. \\
17 & Recover a blue guest folder fully occluded by a larger brown cover. & Cover physically moved first; folder then visible, directly grabbed, and placed on nightstand. \\
18 & Recall the red cola left on a dining table while blue water was refrigerated. & Red cola, not blue water, retrieved and put in opened fridge. \\
19 & Retrieve a label-defined \textsc{Spiced} bottle from a lidded box while a \textsc{White} bottle is visible. & Lid removed; SPICED bottle independently extracted and placed beside WHITE. \\
20 & Retrieve black and blue squeegees nested in a box/cabinet while a white one is a distractor. & Cabinet and box opened, lid removed, both targets independently extracted; black and blue placed on their respective result tables. \\
\end{longtable}
\end{scriptsize}

\section{Evaluation Setting Details and More Experiment Results}
\label{app:experiment}

\subsection{Experiment setting detail of error type analysis.}
\label{app_error}

We developed the error taxonomy from the model results rather than defining it in advance.  Each task family had its own success criteria. These criteria checked the final environment state and, when needed,
  the required action order or intervention time. For every valid failed episode, we reviewed the full conversation,
  visual frames, action results, and final state. We then found the earliest mistake that meaningfully caused the
  failure. Runs caused mainly by missing outputs, backend errors, or simulator problems were not treated as model
  errors.

We first wrote short descriptions of the repeated failure patterns and then merged similar patterns. This produced six
  main categories. R1 means that the model misreads a visible object or location. R2 means that it loses track of what
  happened or ignores an action result. R3 means that it applies the wrong rule or misses an important consequence. R4
  means that it creates or updates the wrong multi-step plan. R5 means that it gives an instruction that cannot be
  executed in the current situation. R6 means that it speaks at the wrong time or fails to speak when help is needed. 

Each failed episode received one primary error label so that it was counted only once in the distribution plots. Other
  problems could be recorded as secondary notes, but they did not change the primary count. Ambiguous cases were checked
  again using the original screenshots, model outputs, action feedback, and evaluator results. Therefore, the pie charts
  show the percentage of model-attributable failures assigned to each error type, rather than the percentage of all
  benchmark episodes.

\subsection{Experiment setting detail of Memory Augmentation.}
\label{app_memory}
We added a simple long-term memory to the assistant. At every step, the assistant produced both its next action and a new memory entry. The new entry was added to all earlier entries, rather than replacing them. The full memory was then included in the input for the next step.

The prompt asked the assistant to make each memory as detailed as possible. It was told to record useful facts such as object names, appearance, locations, states, spatial relations, visible text, and OCR results. It could also record earlier actions and their outcomes. This helped the assistant keep information that might be needed later.

We kept the visual input exactly the same as in the original setting. The assistant received at most 32 images at each step. When more than 32 frames are available, we retain the 8 most recent
frames and uniformly sample 24 from the older history. Therefore, memory was the only change between the original and memory.

\begin{bprompt}{Evaluated Assistant: system prompt with self-managed memory}
  \ttfamily
  You are an egocentric AI assistant helping a person through smart glasses.\\
  You receive the person's spoken request and first-person video in chronological
  order. After every environment action, you receive the new visual segment and
  any sentence the person said.\\[2pt]
  You cannot see the evaluator's score, hidden completion state, answer key, or
  simulator ground truth. Judge progress only from the person's speech and the
  first-person video you actually receive.\\[2pt]
  Choose exactly one response mode each turn:\\
  -- \$silent\$ when speaking now would not help.\\
  -- \$interrupt\$<one concise, natural utterance> for an answer, useful next
  step, clarification, or timely instruction.\\
  -- \$error\$<one concise, natural correction> when the person is currently
  making a mistake.\\
  -- \$done\$ only when you judge the requested task is complete. The world is
  scored immediately when you declare done, and nothing afterwards counts.
  \$done\$ must be the entire response: never attach an answer or instruction to
  it. Speak first with \$interrupt\$, observe the result, and declare done on a
  later turn. Completion requires positive evidence that the person's entire
  requested activity has finished. A quiet moment, harmless action, unchanged
  frame, or absence of visible error is not completion; use \$silent\$ and keep
  watching instead.\\[2pt]
  Use ordinary English. Refer to objects by visible description and location,
  never simulator ids. Do not expose hidden reasoning. Do not claim that an
  unseen or undecidable fact is certain.\\[4pt]
  SELF-MANAGED APPEND-ONLY MEMORY MODE. Maintain a chronological private memory
  ledger across turns. Each request contains at most 32 images selected from the
  visual history. The two latest visual turns are kept intact, while older frames
  are uniformly sampled. Some earlier frames may therefore disappear.\\[2pt]
  The output-format instructions below replace only the earlier formatting
  rules. The meanings of silent, interrupt, error, and done stay unchanged.
  Return exactly these two XML elements in this order, with no text outside them
  and no Markdown fence:\\
  <response>one valid control response: \$silent\$, \$interrupt\$...,
  \$error\$..., or \$done\$</response>\\
  <memory>only the complete new factual information learned on this turn, or
  empty</memory>\\[2pt]
  Memory is private and is never spoken to the person. The system appends each
  non-empty update to the chronological memory ledger. Never write your own
  ``[Memory entry N]'' label. Do not repeat or rewrite facts that are already in
  the memory. If this turn adds no useful new fact or correction, return an empty
  <memory></memory>.\\[2pt]
  Make each new memory entry as detailed as possible. Record task-relevant object
  categories, colors, materials, shapes, sizes, counts, and distinctive marks.
  Record every visible label or OCR string exactly as shown, including capital
  letters, numbers, and units. Use [ILLEGIBLE] or [OCCLUDED] instead of guessing.
  Also record object and person states, precise locations, events, completed
  actions, observed results, unresolved questions, and required follow-up.\\[2pt]
  Use these sections only when they contain new information:\\
  EVENT: actor | action | source -> destination | directly observed result\\
  CURRENT\_STATE: object | state | current location | evidence\\
  HELD: actor | object or empty | evidence\\
  PENDING: unresolved required follow-through | last confirmed step | next check\\
  UNCERTAINTY: claim | alternatives | evidence needed\\[2pt]
  Use short lines and omit unused sections. When a pending item is resolved or an
  object state changes, add an explicit resolution or correction. The newest
  evidence wins. Keep the location of an event separate from the object's current
  location.\\[2pt]
  One action proves only its own result. Walking proves a location change, and
  grabbing proves possession. Neither proves placement, closing, turning
  something off, or task completion. A stated plan or destination also does not
  prove that the action happened. Prefer evidence in this order: a successful
  action result or direct post-action view; a current explicit statement from the
  person; older visual evidence or memory; and inference. Mark conflicts and
  uncertainty instead of treating weak evidence as certain.\\[2pt]
  Before returning \$done\$, check every PENDING item against later evidence.
  Every required physical change must have a successful action result or direct
  post-action evidence. Confirm that no task object remains held unless the
  request requires it. Restore any door, lid, or other object moved for access
  unless the request requires a different final state. If any required placement,
  closing, turn-off, release, or restoration is not confirmed, keep it PENDING
  and do not return \$done\$.\\[2pt]
  Do not put control tokens, hidden reasoning, simulator ids, answer keys, or
  instructions to the person inside memory. The complete accumulated memory
  ledger is provided again on the next turn and is never truncated by the system.
  \end{bprompt}

\subsection{Additional Experiments about Comparative of Dynamic vs. Static}
\label{app:static-benchmark}

We made a static version of the same 120 tasks used in the interactive
benchmark. The final benchmark has six task families: Guidance, Planning,
Accessibility, Decision, Prevention, and Finding. Each family has 20 tasks.
We used the same scenes and recorded first-person episodes as the interactive
benchmark, so we did not create easier scenes for the static test. We stopped
each episode just before the assistant had to answer or make a decision. For a
Prevention task, we stopped after the warning cue appeared but before the
assistant could intervene.

The static model saw a fixed sequence of images instead of a live environment.
Each image sequence was paired with one short question. The question kept the
main goal of the original dynamic task. A Finding item asked where an object
was. A Guidance or Accessibility item asked for the next useful instruction. A
Decision item asked for the answer or choice needed at that point. A Prevention
item asked whether the user should be stopped and what was wrong. A Planning
item asked for the full plan. We wrote the reference answer from the original
task definition and the correct reference run, but only used facts available
before the cut. The model did not see the reference answer, scoring rule,
future frames, action results, or later user feedback. It also had no chance to
act, check the result, or recover from an error.

We evaluated eight models once on every static item. A frozen semantic judge
compared each response with the reference answer. Exact wording was not needed;
the answer passed when it gave the same key fact, location, instruction,
warning, or plan without a harmful contradiction.

\begin{table*}[h]
  \centering
  \caption{Static accuracy across the six task families. Every entry uses a
  fixed denominator of 20 tasks. Static items without enough evidence are
  counted as failures.}
  \label{tab:static-benchmark-results}
  \small
  \setlength{\tabcolsep}{4pt}
  \renewcommand{\arraystretch}{1.12}
  \begin{tabular*}{\textwidth}{@{\extracolsep{\fill}}lcccccc}
    \toprule
    Model & Guidance & Planning & Accessibility & Decision & Prevention & Finding \\
    \midrule
    GPT-5.5                    & 45\% & 50\% & 50\% & 75\% & 70\% & 55\% \\
    GPT-5.6 Sol                & 45\% & 50\% & 60\% & 70\% & 75\% & 50\% \\
    GPT-5.6 Luna               & 35\% & 50\% & 55\% & 60\% & 65\% & 40\% \\
    Claude Opus 5              & 40\% & 45\% & 50\% & 70\% & 70\% & 50\% \\
    Claude Fable 5.1           & 45\% & 65\% & 65\% & 70\% & 75\% & 50\% \\
    Claude Opus 4.8            & 35\% & 30\% & 25\% & 60\% & 45\% & 40\% \\
    Gemini 3.1 Pro    & 60\% & 50\% & 55\% & 50\% & 65\% & 30\% \\
    Muse Spark 1.1             & 45\% & 45\% & 60\% & 65\% & 75\% & 60\% \\
    \bottomrule
  \end{tabular*}
\end{table*}

\begin{figure}[h]
    \centering
    \includegraphics[width=0.5\linewidth]{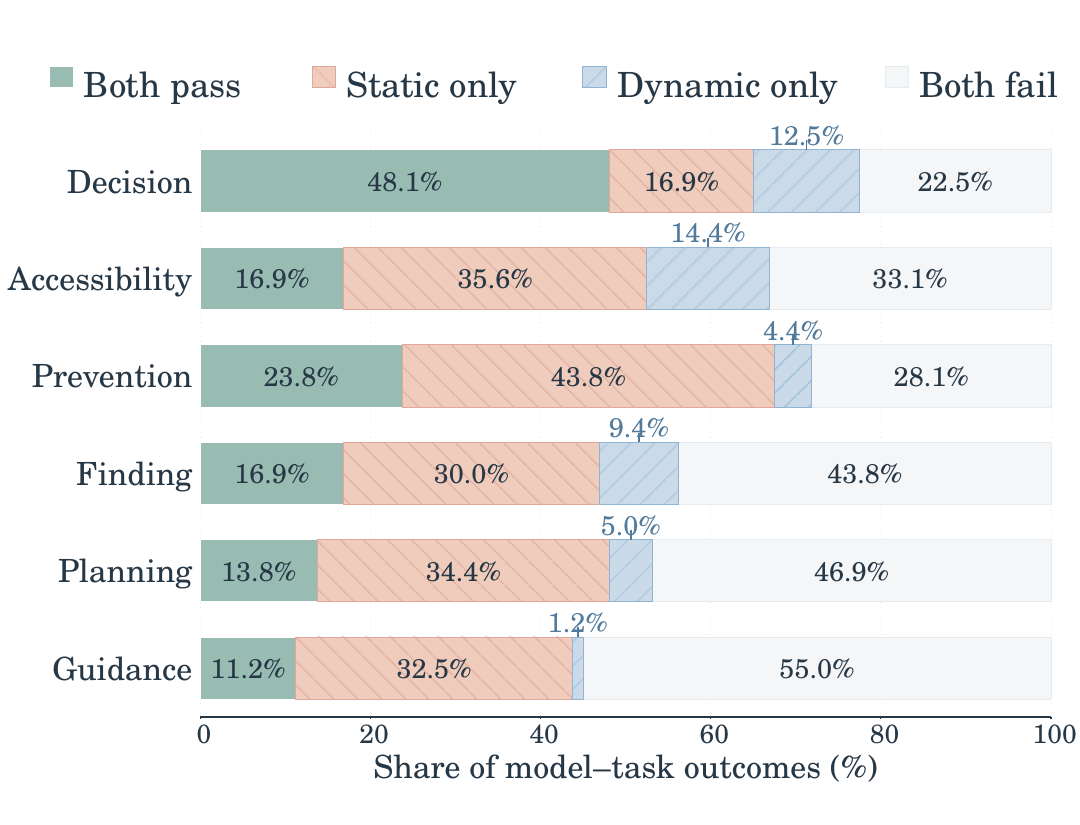}

    \caption{
       \textbf{Performance on Static Adaptations of InteractiveBench} inflate 8 models’ scores
 ; large Static Only (red) and Dynamic Only (blue) regions show a severe evaluation
gap. 
    }
    \label{fig:static}
\end{figure}

\paragraph{Examples of the dynamic-to-static conversion.}
The following examples show how we kept the task goal but removed interaction.
\begin{itemize}
  \item \textbf{Decision (ASK-13011).} The dynamic task asks the assistant to
  read a price sheet and approve a milk order. The static item shows all frames
  before the question and asks for the total. The reference answer is \$41:
  five VAN cartons at \$3, four CHOC cartons at \$4, and two ORIG cartons at
  \$5.

  \item \textbf{Finding (FIND-14001).} The dynamic task asks the assistant to
  find and bring two coconut-milk bottles. The static item asks where the two
  bottles are. The reference answer says that one is in the kitchen
  refrigerator and the other is on the outdoor table in the garden.

  \item \textbf{Guidance (GUIDE-90002).} The dynamic task asks the assistant to
  guide the user through preparing an area for visitors. The static item keeps
  the viewed checklist and asks what must happen first. The reference answer is
  to close the open green staff locker because the checklist puts privacy
  first.

  \item \textbf{Accessibility (NARRATE-87016).} The static input shows the
  cleanup guide, broken glass, and nearby tools. It asks for the next useful
  instruction. The reference answer says to handle the broken glass first and
  place the shard in the open yellow paper bag, without using bare hands.

  \item \textbf{Planning (ORCHESTRATE-14008).} The dynamic task asks the
  assistant to guide a multi-room move of a folder, a phone, and a mug. The
  static item asks for the complete efficient plan. The reference route is:
  move the folder to the bed, take the phone from that bed to the dining table,
  walk empty to the other living room, and move the mug to the kitchen counter.

  \item \textbf{Prevention (WATCH-16005).} The user says that the dough is
  ready and will go straight into the oven. The static item asks whether the
  assistant should interrupt. The reference answer is yes: score a shallow X
  in the dough first, return the knife, then put the dough in the oven and close
  the door.
\end{itemize}

\subsection{Additional Experiment of Task Horizon and Model Performance}
  \label{sec:horizon_analysis}

  In this experiment, we asked a simple question: do models succeed less often when a task requires a longer sequence of actions? We used 120
  tasks from six task families. We evaluated eight models on each task.
  We measured task length using a reference solution, rather than the number of actions taken by the model. This avoids a
  major bias: a failed model may repeat actions or run until the step limit, while a successful model may stop early. Let
  \(B_t\) be the best available reference action count for task \(t\). We normalized this count separately within each
  task family:
  \[
  h_t =
  \frac{
  B_t-\min_{j\in\mathcal{F}(t)} B_j
  }{
  \max_{j\in\mathcal{F}(t)} B_j
  -\min_{j\in\mathcal{F}(t)} B_j
  },
  \]
  where \(\mathcal{F}(t)\) is the family containing task \(t\). Thus, \(h_t=0\) represents the shortest reference task in
  that family, and \(h_t=1\) represents the longest. These values are relative to each family. The horizontal axis
  compares different tasks; it does not show progress within one episode.

\begin{figure*}[t]
      \centering
      \includegraphics[width=\textwidth]{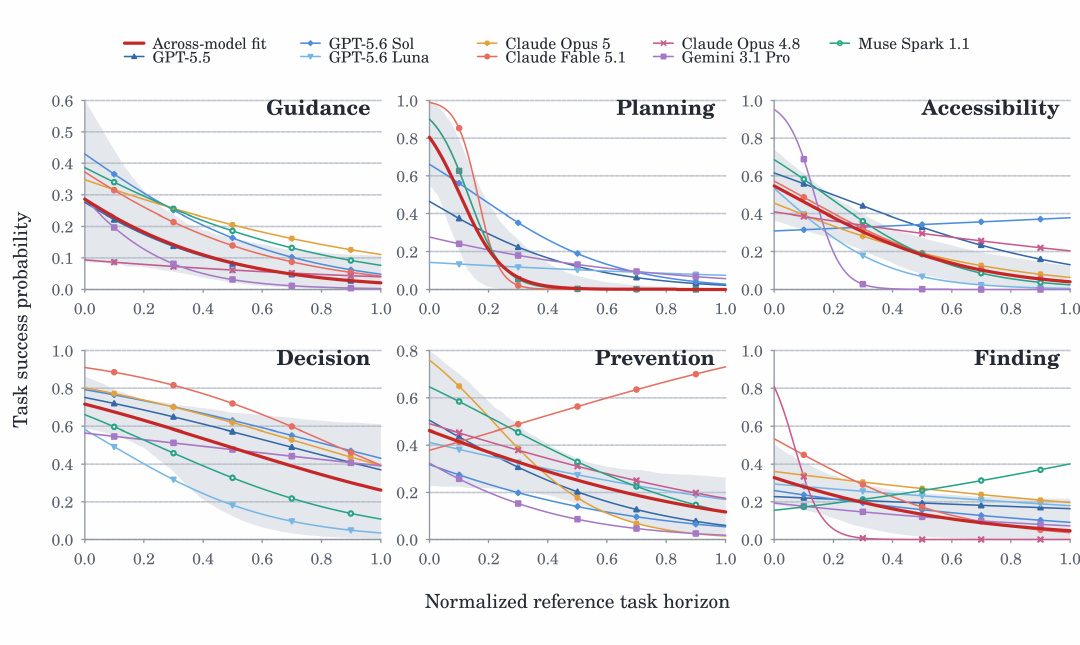}
      \caption{
      Complete-task success probability versus normalized reference task
      horizon across six task families. Each colored curve is a
      model-specific Firth logistic fit. The red curve is the pooled fit
      across models, and the gray band is its pointwise \(95\%\)
      task-bootstrap interval. Horizon is normalized separately within each task family.
      The figure compares different tasks and does not represent progress
      within a single episode. No monotonicity constraint was applied.
      }
      \label{fig:horizon_success_all_families}
  \end{figure*}

  For visualization, we fitted a separate Firth logistic regression for every model and task family. The colored curves in
  Figure~\ref{fig:horizon_success_all_families} show the fitted success probability of each model. The red curve is the
  pooled fit across models, and the gray band is its pointwise \(95\%\) task-bootstrap interval. The curves were not
  forced to decrease. For the main statistical result, we first averaged the eight binary model outcomes for each task and
  then fitted a task-level linear probability model. The overall model included task-family fixed effects. During
  bootstrap and permutation tests, all eight outcomes from the same task were kept together.

  The family-adjusted slope was \(-0.431\), corresponding to a \(43.1\)-percentage-point decrease from the shortest to the
  longest normalized horizon (\(95\%\) bootstrap CI: \(-58.1\) to \(-28.9\); one-sided permutation \(p<0.0001\)). All six
  family-level point estimates were negative, and 45 of the 48 model--family curves also decreased. However, not every
  family-level test was individually significant. These results show an association between longer tasks and lower
  complete-task success, but they do not prove that task length directly causes failure. Longer tasks may also require
  more search, more object interactions, more state changes, or stricter safety constraints. Some labels were
  reconstructed from saved traces, so we treat this result as provisional.

\subsection{Additional Human-Agent Validation with Real Human Participants}
\label{app:human_validation}

\paragraph{Purpose and setup.}
We examine whether replacing a model-based Human agent with a human
participant changes the evaluation of an Assistant. We compare four
model-based executors---Fable 5.1, Muse 1.1, GPT-5.6, and Gemini 3.1---with
three human participants in InteractiveWorld. The comparison covers the
same 36 tasks, with six tasks from each of the six benchmark families.
Each executor is paired with each of four Assistants. The human
participants take the executor role in the simulator; this is not a
physical-world deployment study. We report the number of fully completed
tasks out of 36 for each Assistant--executor pairing.

The model-based Human agents follow the instruction-following protocol
specified in Appendix~C.2.4. This prompt asks them to carry out the
Assistant's guidance, ground object references before acting, and request
clarification when an instruction is ambiguous. The comparison therefore
concerns prompted agents in this defined executor role, rather than
unrestricted agents pursuing their goals.

\begin{table}[htbp]
    \centering
    \caption{
        \textbf{Assistant performance with model-based and human
        executors.}
        Rows identify the evaluated Assistant; columns identify the
        executor. Each entry is the number of successful tasks out of
        the same 36-task subset. Human 1--3 are three individual
        humans, not three independent task sets.
    }
    \label{tab:human_executor_validation}

    \small
    \setlength{\tabcolsep}{4pt}
    \renewcommand{\arraystretch}{1.15}

    \resizebox{\linewidth}{!}{%
    \begin{tabular}{lccccccc}
        \toprule
        & \multicolumn{4}{c}{Model-based Human agent}
        & \multicolumn{3}{c}{Human participant} \\
        \cmidrule(lr){2-5}
        \cmidrule(lr){6-8}

        Assistant
        & Fable 5.1
        & Muse 1.1
        & GPT-5.6
        & Gemini 3.1
        & Human 1
        & Human 2
        & Human 3 \\

        \midrule
        Fable 5.1
        & 13/36 & 15/36 & 13/36 & 13/36
        & 15/36 & 16/36 & 15/36 \\

        Muse 1.1
        & 9/36 & 9/36 & 8/36 & 9/36
        & 10/36 & 11/36 & 9/36 \\

        GPT-5.6 Sol
        & 8/36 & 9/36 & 11/36 & 10/36
        & 8/36 & 11/36 & 9/36 \\

        Gemini 3.1
        & 8/36 & 8/36 & 5/36 & 7/36
        & 8/36 & 8/36 & 7/36 \\
        \bottomrule
    \end{tabular}
    }

    \vspace{3pt}
    \begin{minipage}{\linewidth}
    
    \end{minipage}
\end{table}

\begin{figure}[h]
    \centering
    \includegraphics[width=\linewidth]{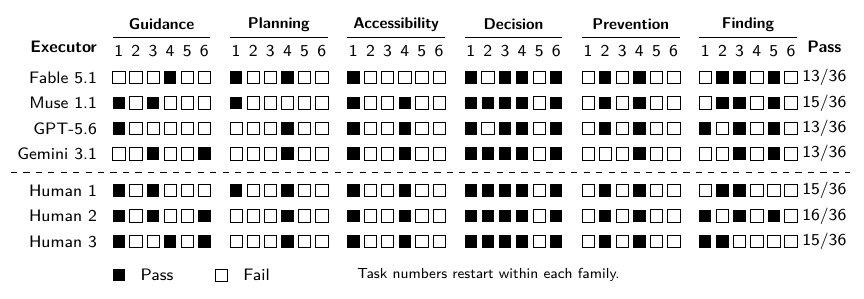}

    \caption{
        \textbf{Task-level outcomes for Fable 5.1 with different
        executors.}
        Rows are executors and columns are the 36 tasks, grouped into
        six families. Filled squares indicate success and open squares
        indicate failure. Task numbers restart within each family.
        The dashed line separates model-based Human agents from human
        participants.
    }
    \label{fig:fable_human_outcomes}
\end{figure}
\paragraph{Executor changes preserve the broad performance pattern.}
Table~\ref{tab:human_executor_validation} shows that Fable 5.1 is the
best-performing Assistant with every executor, while Gemini 3.1 is
lowest or tied for lowest. Switching among the four model-based Human
agents changes each Assistant's pass count by at most three tasks out
of 36, or 8.3 percentage points. Across all seven executors, the largest
observed range is three tasks, or 8.3 percentage points. Executor choice
therefore affects individual scores, but does not overturn the broad
performance pattern in this subset. The ordering of Muse 1.1 and
GPT-5.6 Sol does change across executors, so we do not claim that the
complete ranking is invariant.

\paragraph{Pass rates are close to those obtained with human participants.}
For Gemini 3.1, the Human agent used in our main evaluation, the four
Assistants achieve pass rates of 36.1\%, 25.0\%, 27.8\%, and 19.4\%,
in the row order of Table~\ref{tab:human_executor_validation}. Averaging
over the three human participants gives 42.6\%, 27.8\%, 25.9\%, and
21.3\%, respectively. The human-minus-agent differences are therefore
$+6.5$, $+2.8$, $-1.9$, and $+1.9$ percentage points. The observed
scores are therefore relatively close, although the executor-related
differences are not zero.

\paragraph{Agreement extends to individual task outcomes.}
To check whether similar overall pass rates hide different successful
tasks, we inspect the 36 task-level outcomes for Fable 5.1 as the
Assistant (Figure~\ref{fig:fable_human_outcomes}). The Gemini 3.1
executor agrees with Human 1, Human 2, and Human 3 on 30/36, 33/36,
and 28/36 tasks, respectively. Their mean outcome agreement is 84.3\%.
For comparison, the three human--human pairs agree on 31/36, 30/36,
and 31/36 tasks, giving a mean of 85.2\%.

Thus, for this Assistant, agent--human agreement is close to the
agreement observed between the human participants themselves.
Agreement counts both shared successes and shared failures; it measures
consistency of task outcomes, not identical actions or dialogue.
These descriptive comparisons reuse the same tasks and are not
independent samples or a statistical equivalence test.

\paragraph{Implications and scope.}
Together, these results provide preliminary support for prompted Human
agents as a practical proxy for human instruction following in
InteractiveWorld. Both the broad Assistant comparisons and the detailed
Fable 5.1 outcomes show correspondence with evaluations involving real
participants. This addresses whether the benchmark results depend
entirely on a particular simulated executor.

\section{Future Work}
\label{app:futurework}
During this study, we identified several interesting directions to explore next. (1) Since InteractiveWorld provides verifiable, state-based rewards, we will build a reinforcement learning platform on it to train interactive abilities efficiently. (2) We will make the Human agent behave more like real people, through training or other means, and simulate users with different personalities and skill levels. (3) As InteractiveWorld supports scanned real-world scenes, we will bring in more real environments and situations where people need an assistant, making the tasks more realistic.

\end{document}